\documentclass[9pt,twocolumn,twoside]{pnas-new}

\templatetype{pnasresearcharticle} 

\usepackage[normalem]{ulem}
\usepackage{pdfpages}

\usepackage{fancyhdr} 
\fancypagestyle{firststyle}{
  \fancyhead[LO]{}
  \fancyhead[RO]{}
  \fancyhead[LE]{}
  \fancyhead[RE]{}
}

\begin{document}

\title{A Precise Metallicity and Carbon-to-Oxygen Ratio for a Warm Giant Exoplanet from its Panchromatic JWST Emission Spectrum}

\author[a, 1]{Lindsey S. Wiser}
\author[b,c,d]{Taylor J. Bell}
\author[a]{Michael R. Line}
\author[e]{Everett Schlawin}
\author[f]{Thomas G. Beatty}
\author[a]{Luis Welbanks}
\author[c]{Thomas P. Greene}
\author[h]{Vivien Parmentier}
\author[e]{Matthew M. Murphy}
\author[g]{Jonathan J. Fortney}
\author[f]{Kenny Arnold}
\author[h]{Nishil Mehta}
\author[i]{Kazumasa Ohno}
\author[g]{Sagnick Mukherjee}

\affil[a]{School of Earth \& Space Exploration, Arizona State University, Tempe, AZ 85287}
\affil[b]{Bay Area Environmental Research Institute, NASA Ames Research Center, Moffett Field, CA 94035}
\affil[c]{Space Science and Astrobiology Division, NASA Ames Research Center, Moffett Field, CA 94035}
\affil[d]{AURA for the European Space Agency, Space Telescope Science Institute, Baltimore, MD 21218}
\affil[e]{Department of Astronomy, Steward Observatory, University of Arizona, Tucson, AZ 85721}
\affil[f]{Department of Astronomy, University of Wisconsin--Madison, Madison, WI 53706}
\affil[g]{Department of Astronomy and Astrophysics, University of California, Santa Cruz, CA 95064}
\affil[h]{Université Côte d'Azur, Observatoire de la Côte d'Azur, CNRS, Laboratoire Lagrange, France}
\affil[i]{Division of Science, National Astronomical Observatory of Japan, Tokyo, 181-8588, Japan}

\leadauthor{Wiser}

\significancestatement{Over the past three decades, exoplanet research has progressed from planet discovery to detecting molecules in exoplanet atmospheres and identifying population-level trends that inform hypotheses about planet formation. Hot gas giants around low-mass stars are relatively rare, challenging our understanding of how planets form. In this paper, we consider the composition of one of these rare planets, WASP-80 b. We present observations from the revolutionary James Webb Space Telescope, and we discuss how this unique planet may have formed.}

\authorcontributions{L.S.W., T.J.B., M.R.L., L.W., and T.P.G. designed research; L.S.W., T.J.B., M.R.L., E.S., T.G.B., L.W., T.P.G., V.P., M.M.M., J.J.F., K.A., N.M., K.O., and S.M. performed research; L.S.W., T.J.B., E.S., and T.G.B. analyzed data; and L.S.W., T.J.B., E.S., and T.G.B. wrote the paper.}
\correspondingauthor{\textsuperscript{1}To whom correspondence should be addressed. \\E-mail: lindsey.wiser@asu.edu}


\begin{abstract}
WASP-80~b, a warm sub-Jovian (equilibrium temperature $\sim$820~K, 0.5~Jupiter masses), presents an opportunity to characterize a rare gas giant exoplanet around a low-mass star. In addition, its moderate temperature enables its atmosphere to host a range of carbon and oxygen species (H$_2$O, CH$_4$, CO, CO$_2$, NH$_3$). In this paper, we present a panchromatic emission spectrum of WASP-80 b, the first gas giant around a late K/early M-dwarf star and the coolest planet for which the James Webb Space Telescope has obtained a complete emission spectrum spanning 2.4--12~$\mathrm{\mu m}$, including NIRCam F322W2 (2.4--4~$\mathrm{\mu m}$) and F444W (4--5~$\mathrm{\mu m}$), and MIRI LRS (5--12~$\mathrm{\mu m}$). We report confident detections of H$_2$O, CH$_4$, CO, and CO$_2$, and a tentative detection of NH$_3$. We estimate WASP-80~b’s atmospheric metallicity and carbon-to-oxygen ratio and compare them with estimates for other gas giants. Despite the relative rarity of giant planets around low-mass stars, we find that WASP-80~b’s composition is consistent with other hot gas giants, suggesting that the formation pathway of WASP-80~b may not be dissimilar from hot gas giants around higher-mass stars. 
\end{abstract}

\dates{This manuscript was compiled on \today. In press with PNAS as of March 2025.}
\doi{\url{www.pnas.org/cgi/doi/10.1073/pnas.XXXXXXXXXX}}

\maketitle
\thispagestyle{firststyle}
\ifthenelse{\boolean{shortarticle}}{\ifthenelse{\boolean{singlecolumn}}{\abscontentformatted}{\abscontent}}{}

\firstpage[11]{3}

\dropcap{T}he 2021 launch of the James Webb Space Telescope (JWST) marked the start of a new era in exoplanet characterization. Over the past three decades, our understanding of exoplanets has progressed from planet discovery to the detection of molecules in their atmospheres and even the beginning of population-level trends that inform hypotheses about planet formation and the unique climates of distant worlds \cite[e.g., ][]{Goyal2021, Changeat2022, Wiser2024}. The sensitivity, wavelength coverage, and spectral resolution of JWST enable the characterization of a greater variety of exoatmospheric properties with increased confidence \cite[e.g., ][]{Taylor2023, Grant2023, Bell2023, Bean2023, Welbanks2024, Murphy2024arXiv, Bell2024W43, Nixon2024, Schlawin2024, Beatty2024}, thus informing our predictions about compositions, chemical processes, climates, and planet formation, as well as broadening our picture of exoplanets as a whole. 

\begin{figure}
    \centering
    \includegraphics[width=1.0\linewidth]{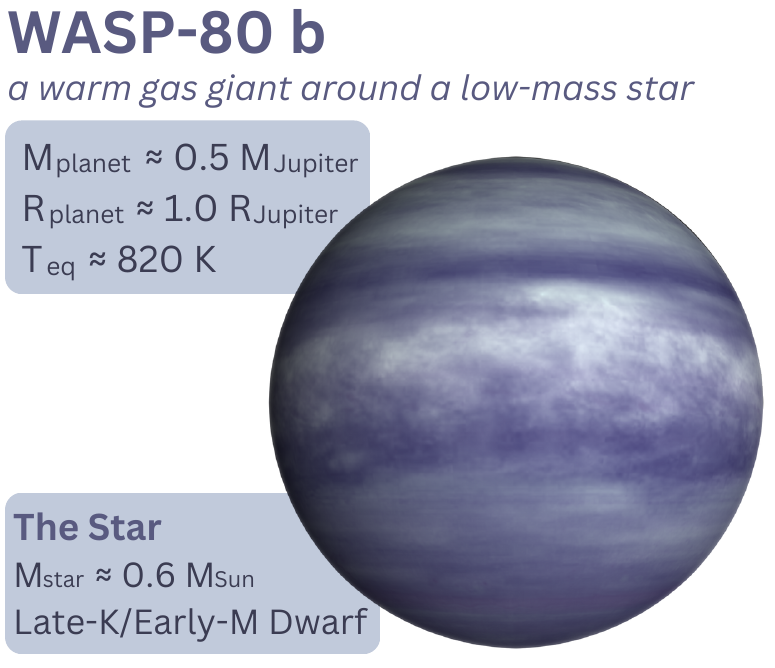}
    \caption{An artist's rendering of the warm exoplanet WASP-80 b and an overview of its planet-system parameters. Planet image credit: NASA/Ames Research Center.}
    \label{fig:art}
\end{figure}

Revealing the nature of exoplanets provides context for our own solar system. An overarching feature of the growing population of known exoplanets is that many planets are unlike those in our solar system, challenging our understanding of planet formation \cite[e.g.,][]{IdaLin2004, Fortney2021}. One such planet is WASP-80 b (see Figure \ref{fig:art} for an artist rendering). WASP-80~b ($0.5~M_{\rm Jupiter}$, $1.0~R_{\rm Jupiter}$) has been described as the ``missing link'' \cite{Triaud2015} in our understanding of planet formation and atmospheric physics. This is due to its sub-Jovian mass, its moderate temperature ($T_\mathrm{eq.}\approx820~$K) that supports the existence of multiple carbon species (CH$_4$, CO, and CO$_2$), and its close proximity to its host star (a late K/early M-dwarf, $0.6~M_{\rm Sun}$, orbital semi-major axis~=~0.034 AU), making it susceptible to atmospheric photochemistry \cite[e.g.,][]{Moses2011, Tsai2017}. Further, WASP-80~b is the only planet in this temperature, mass, and host-star regime with a near complete broadband thermal emission spectrum 2.4 to 12~$\mu$m \cite{Fukui2014, Wong2022, Bell2023}. 

\begin{figure}
    \centering
    \includegraphics[width=1.0\linewidth]{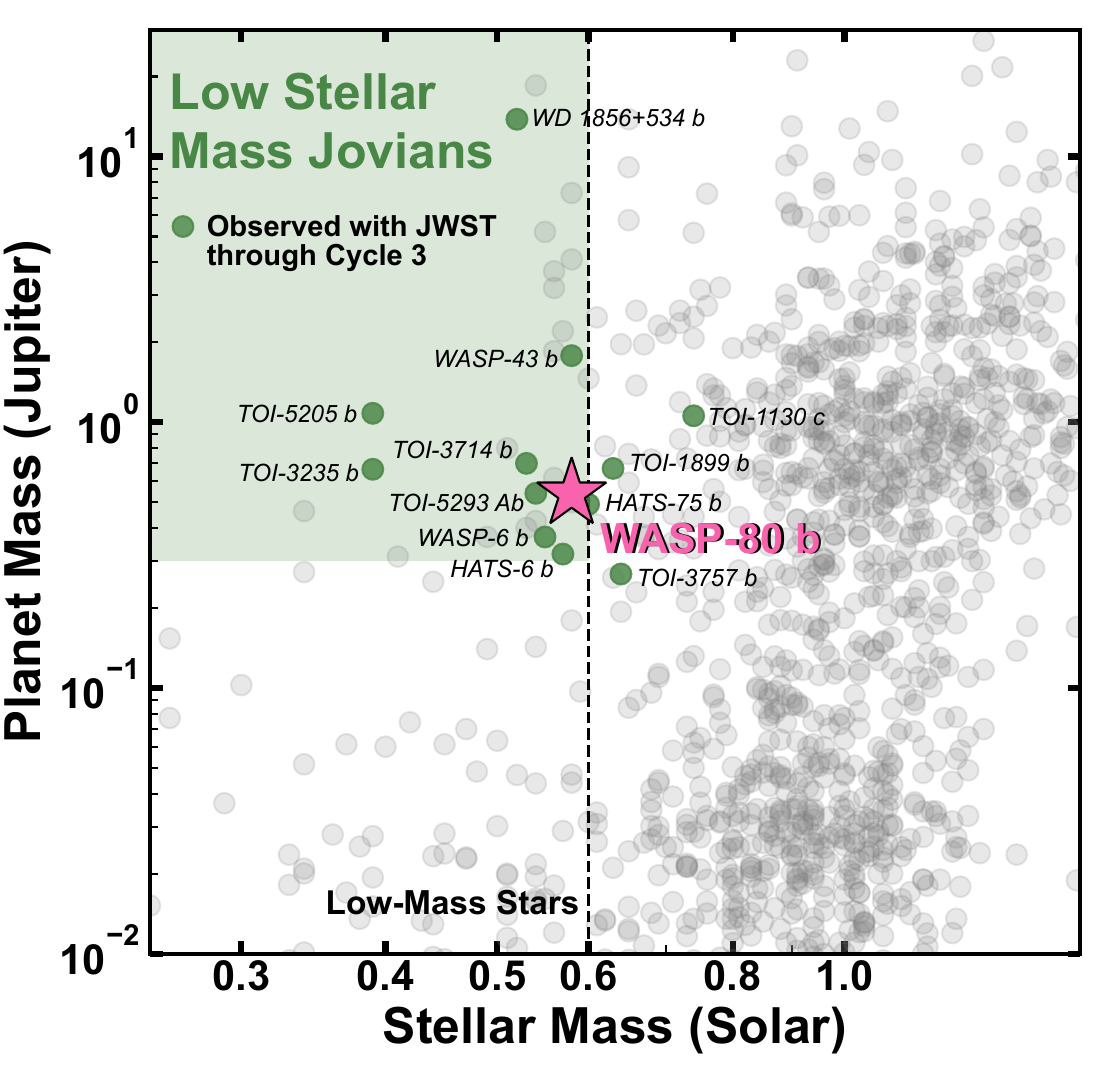}
    \caption{Transiting giant planets around low-mass stars that have been observed, or will be observed, with JWST through Observation Cycle 3. We define these ``low stellar mass Jovians" as planets with stellar mass $<0.6 M_{\rm Sun}$ and planet mass $>0.3 M_{\rm Jupiter}$ and shade that region in green. JWST targets with masses just outside of those ranges are also plotted in green. Grey points show all other known transiting planets from the NASA Exoplanet Archive.}
    \label{fig:population}
\end{figure}

Planet population statistics indicate that, in general, massive planets tend to exist around massive stars \cite{Adams2005}. In Figure \ref{fig:population}, we highlight the rare population of massive planets around low-mass stars, ``low stellar mass Jovians." Following the core accretion hypothesis for planet formation, grains in a protoplanetary disk first form a planet's core, followed by the accretion of additional disk material to build up a massive atmosphere \cite{Pollack1996, Chabrier2014}. However, around a low-mass star, $<$~0.6 $M_{\rm Sun}$, the protoplanetary disk may not contain enough mass to easily build Jovian-sized planets \cite{Adams2005, Burn2021}. WASP-80 b, therefore, offers a unique opportunity to characterize a massive planet orbiting a host star on the edge of this low-mass star cutoff and to assess whether WASP-80 b's composition is consistent with other gas giant atmospheres. We seek precise constraints on WASP-80 b's atmospheric metallicity ([M/H], in which M includes all non-H/He elements and [] denotes log$_{10}$ relative to solar abundances) and carbon-to-oxygen ratio (C/O). [M/H] and C/O are commonly estimated metrics hypothesized to hint at a planet's formation location and disk migration because of varying gas and solid/ice compositions at different orbital radii in a protoplanetary disk \cite{Oberg2011, Mordasini2016, Batygin2016, Madhu2017, Booth2017, Reggiani2022, Chachan2023}. 

WASP-80~b's warm temperature is thermochemically conducive to numerous species, including CH$_4$, CO, CO$_2$, NH$_3$, and H$_2$O \cite{Schlawin2024}. Ref.\ \cite{Bell2023} reported the detection of H$_2$O and previously elusive CH$_4$ in WASP-80~b's atmosphere from JWST NIRCam F322W2 observations in both transmission and emission. Emission observations, in particular, are valuable for sampling the full dayside of a planet, rather than just the limbs. Emission observations are less impacted by stellar activity that can contaminate transmission observations; WASP-80, being a late K/early M-dwarf, is a likely active star \cite{Mancini2014, Triaud2013}. In this paper, with the addition of NIRCam F444W and MIRI LRS emission observations, we increase confidence in detections of CH$_4$ and H$_2$O, we add detections of CO, CO$_2$, and possible NH$_3$, and we refine estimates for the atmospheric [M/H] and C/O. We conclude by exploring what WASP-80 b's atmospheric composition reveals about its formation, placing these new insights within the context of JWST's impact on exoplanetary science. 

\section*{Methods}

\subsection*{Observations and Data Reduction}
We present a panchromatic emission spectrum of WASP-80 b, 2.4--12.0 $\mu$m, collected through the MANATEE JWST Guaranteed Time Observations (GTO) program. Observations include NIRCam F322W2 (taken 2022 Oct 29; JWST-GTO-1185 Observation 4), NIRCam F444W (taken 2023 Jun 13; JWST-GTO-1185 Observation 5), and MIRI LRS (taken 2022 Sep 25; JWST-GTO-1177 Observation 2). The NIRCam F322W2 (2.4--4.0 $\mu$m) emission observations were previously presented in ref.\ \cite{Bell2023}, while this is the first publication of the F444W and LRS data. For the MIRI LRS data, we conducted two independent data reductions using \texttt{Eureka!} and \texttt{tshirt}, and for both of the NIRCam observations, we performed three independent reductions using \texttt{Eureka!}, \texttt{tshirt}, and \texttt{Pegasus}; the raw and fitted lightcurves from the \texttt{Eureka!} reduction are shown in Supporting Information Figure S1. Each reduction and fitting method is described below. As shown in Figure \ref{fig:data}, all reductions were quite consistent, with a mean difference and mean absolute difference between \texttt{Eureka!} and \texttt{tshirt} of only 0.2$\sigma$ and 0.4$\sigma$, respectively. Between \texttt{Eureka!} and \texttt{Pegasus}, the mean difference and mean absolute difference are -0.03$\sigma$ and 0.4$\sigma$, respectively. We ultimately selected the \texttt{Eureka!} reduction as our fiducial emission spectrum for the modeling analysis since it was the most performant pipeline across the entire wavelength range (see Supporting Information Figures S2--S3) and provided a uniform panchromatic spectrum, while also being generally consistent with multiple independent reductions. Comparing the \texttt{Eureka!} reduction to the reduction of F322W2 previously published in ref. \cite{Bell2023}, the reductions agree within 0.52$\sigma$.

\subsubsection*{Eureka!\ Reduction (Fiducial)}
Our fiducial reduction used the open-source \texttt{Eureka!} package \cite{Bell2022} with version 0.11.dev51+gbddf46c5, version 1.10.2 of the \texttt{jwst} package, and CRDS version 11.17.0 with CRDS context \texttt{jwst\_1100.pmap} for NIRCam F322W2 and F444W and \texttt{jwst\_1105.pmap} for MIRI LRS. The \texttt{Eureka!} control and parameter files we used are available for download\footnote{\url{https://doi.org/10.5281/zenodo.13146949}} and the important parameters are summarized below.

Our analysis closely followed the methods in ref.\ \cite{Bell2023} for the NIRCam data and ref.\ \cite{Bell2024} for the MIRI data. We began with the \texttt{\_uncal.fits} files for all three observations. In Stage 1, the jump step rejection threshold was increased to 6.0$\sigma$ for NIRCam and 8.0$\sigma$ for MIRI, and for MIRI, the firstframe and lastframe steps were also run, which removed the first and last frames, respectively. We did not apply the 390 Hz MIRI/LRS noise correction step presented in ref.\ \cite{Welbanks2024} as it was not found to have a significant impact on our final spectra. In Stage 2, the photom and extract1d steps were turned off as they are not needed for time-series observation data. In Stage 3, we cropped to a smaller subarray of interest, converted the data to electrons (for MIRI, we assumed a fixed gain of 3.1 electrons per Data Number following ref.\ \cite{Bell2024}), rotated the MIRI images so that they were oriented like the NIRCam data, corrected for the slight curvature in the NIRCam spectral trace with integer-pixel movements, and performed background subtraction per `column' (where we define a column to be a set of pixels that span the spatial direction). We then performed optimal spectral extraction \cite{horne1986optspec} using the median frame as a spatial profile. In Stage 4, we masked a small number of wavelengths ($\lesssim$3\%) where the standard deviation of the lightcurve was much higher than neighboring pixels (likely due to unmasked bad pixels in earlier stages), spectrally binned the data, and then sigma-clipped outliers in the temporal direction to remove any residual cosmic rays missed in earlier stages.

\begin{table*}
    \centering
    \begin{tabular}{c|c|c}
        \hline \hline & Prior (ref.\cite{Triaud2015}) & Eureka!'s Joint Broadband Fit Posterior \\ \hline
        $R_{\rm p}/R_*$ (F322W2 broadband) & 0.17 $\pm$ 0.01 & 0.17157 $\pm$ 0.00028 \\ 
        $P$ (days) & 3.06785234 $\pm$ 0.00000083 & 3.067851954 $\pm$ 0.000000030 \\ 
        $t_0$ (BJD$_{\rm TDB}$) & 2456487.425006 $\pm$ 0.000025 & 2456487.425008 $\pm$ 0.000024 \\ 
        $a/R_*$ & 12.63 $\pm$ 0.1 & 12.627 $\pm$ 0.035 \\ 
        $i$ ($^{\circ}$) & 89.02 $\pm$ 0.1 & 88.973 $\pm$ 0.069 \\ 
        $e$ & 0 & 0 \\ \hline \hline
    \end{tabular}
    \caption{Our updated WASP-80 b orbital parameters were used by all pipelines when fitting the spectroscopic eclipse lightcurves. BJD$_{\rm TDB}$ is the date in the Barycentric Julian Date in the Barycentric Dynamical Time system. For a justification of our $e=0$ assumption, see Supporting Information Table S1 and Figure S4.}
    \label{tab:orbitalParameters}
\end{table*}

We improved upon the orbital solution of the planet by performing a joint broadband fit of the NIRCam F322W2 transit and eclipse data first presented by ref.\ \cite{Bell2023} as well as the NIRCam F444W and MIRI LRS data first presented in this paper. Our astrophysical and systematic model setups were the same as those of our spectroscopic fits described below, and the adopted orbital solution from this fit is tabulated in Table \ref{tab:orbitalParameters}. We then fit each spectral channel with a \texttt{starry} \cite{luger2019starry} model for the secondary eclipse signal (assuming a uniform brightness distribution across the planet's dayside). We investigated the ability of these data to constrain the heat distribution across the planet's dayside using the exoplanet eclipse mapping method \cite{williams2006resolving, rauscher2007toward} on the broadband lightcurves from each eclipse observation as has been done with previous JWST eclipse observations \cite[e.g.,][]{Hammond2024ERS, Schlawin2024, Coulombe2023ERS}. However, a Bayesian Information Criterion (BIC) test \cite{Schwarz1978} strongly favored a uniform brightness model for the planet over the inclusion of a simple non-uniform brightness eclipse model and/or orbital phase variations (allowing only first-order terms) with a $\Delta$BIC of $\sim$18 for each of the three observations.

\begin{SCfigure*}[\sidecaptionrelwidth][t]
\centering
\includegraphics[width=1.2\linewidth]{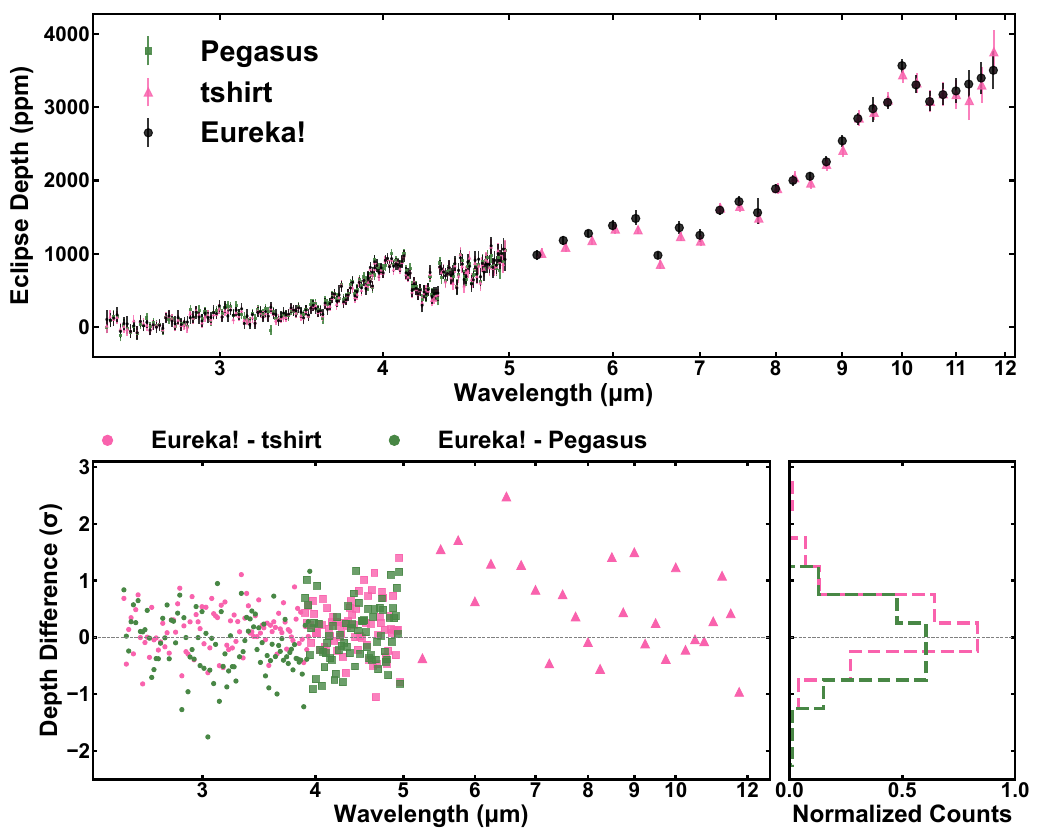}
\caption{\textit{Top:} Our three data reductions -- \texttt{Eureka!}, \texttt{tshirt}, and \texttt{Pegasus}. \texttt{Eureka!} and \texttt{tshirt} were used for NIRCam F322W2, NIRCam F444W, and MIRI LRS, while \texttt{Pegasus} was only used for the two NIRCam filters. \textit{Bottom Left:} Differences between our fiducial \texttt{Eureka!} spectrum and our two supporting reductions for WASP-80 b's emission spectrum. NIRCam F322W2 points are shown with circles, NIRCam F444W points are shown with squares, and MIRI LRS points are shown with upward triangles. \textit{Bottom Right:} A normalized histogram showing the distribution of differences across all detectors. The \texttt{Eureka!}, \texttt{tshirt}, and \texttt{Pegasus} spectra all agree excellently with the vast majority of points lying within -1$\sigma$ and +1$\sigma$, with a very small bias ($\sim$0.2$\sigma$ on average) towards \texttt{tshirt} having smaller eclipse depths than \texttt{Eureka!}.}
\label{fig:data}
\end{SCfigure*}

Our systematic model for each instrument included a linear trend in time, a linear decorrelation against changes in the spatial point spread function (PSF) position and width, a Gaussian Process with a Mat\'ern-3/2 kernel (as implemented by \texttt{celerite2} \cite{celerite1, celerite2}) as a function of time, and a white noise multiplier to account for any additional white noise or errors in the estimated gain used in Stage 3. For the MIRI LRS data, we also included a single exponential ramp following ref.\ \cite{Bell2024}. We removed the first 150 integrations from the NIRCam data (2629 seconds for F322W2 and 5053 seconds for F444W) to ensure the detector had settled, and for the MIRI data we followed ref.\ \cite{Bell2024} by removing the first 800 integrations (3817 seconds) and integrations 3363--3369 which showed a sudden spike in noise. We also tried removing no integrations from the NIRCam data and fewer integrations from the MIRI data, but found that our precision was degraded. We then sampled the posteriors using \texttt{pymc3}'s No U-Turns Sampler \cite{salvatier2016pymc3}, and we ensured that the sampler had converged by requiring that the Gelman-Rubin statistic \cite{GelmanRubin1992} was below 1.1 as computed using a set of three independent chains. Our final \texttt{Eureka!} spectra are shown in Figures \ref{fig:data} and \ref{fig:spec}.

\subsubsection*{tshirt Reduction}
We used the \texttt{tshirt} code (available online\footnote{\url{https://github.com/eas342/tshirt} and \url{https://github.com/eas342/jtow}}) to independently analyze the emission spectrum of WASP-80 b, in a similar manner as in \cite{Bell2023,Welbanks2024,Schlawin2024,Beatty2024}.
We first processed the \texttt{\_uncal.fits} data with \texttt{jwst} Stage 1 version 1.13.4, using CRDS version 11.17.15 and CRDS context \texttt{jwst\_1230.pmap}.
For the NIRCam data, we replaced the reference pixel step with a row-by-row odd/even by amplifier (ROEBA) correction and also set the jump rejection threshold to 6$\sigma$.
We did not apply a ROEBA correction to the MIRI LRS data and used a jump threshold of 7$\sigma$.
We then manually divided the NIRCam \texttt{rateints} data products by imaging reference file flats \texttt{jwst\_nircam\_flat\_0313.fits} for the F444W observation and \texttt{jwst\_nircam\_flat\_0266.fits} available on CRDS.
We traced the spatial location of the source with a Gaussian fit (along the Y/vertical pixel) and then fit the centroid with a sigma-clipped quadratic polynomial as a function of X/horizontal pixel.
We then extracted a 10-pixel full-width aperture for NIRCam and an 8-pixel full-width aperture around the source, rounding to the nearest whole pixel.
Finally, we binned to the wavelength grid to the same as the \texttt{Eureka!} reduction, rounding wavelengths to the nearest whole pixel.

We fit the lightcurves with \texttt{starry} \cite{luger2019starry} and \texttt{pymc3} \cite{salvatier2016pymc3}.
We assumed a uniform dayside when obtaining spectroscopic transit depths.
We fixed the orbital parameters to the joint analysis (described above and tabulated in Table \ref{tab:orbitalParameters}).
When fitting the NIRCam lightcurves, we included a linear trend as a function of time and a linear function in terms of the focal plane array housing temperature (FPAH) as in ref. \cite{Schlawin2024}.
For the MIRI LRS lightcurves, we only included a linear trend with time and an exponential ramp with a prior of a 70-minute settling time.
We discarded the first 700 integrations with the MIRI LRS lightcurve to reduce the settling behavior and did not discard any data for the NIRCam F322W2 or F444W lightcurves.
We used a 5 sigma clipping algorithm for cosmic rays and simultaneously fit for the uncertainty in the lightcurve data to allow for error inflation beyond photon and read noise.

\subsubsection*{Pegasus Reduction}
We also reduced the NIRCam observations using the \texttt{Pegasus} pipeline\footnote{\url{https://github.com/TGBeatty/PegasusProject}}. Details of the \texttt{Pegasus} data reduction are described in detail in ref.\ \cite{Beatty2024}, and we summarize them here. We began by conducting background subtraction on the \texttt{rateint} files provided by version 1.10.2 of the \texttt{jwst} pipeline using CRDS version 11.17.0. We masked the area on the images around the light from WASP-80. Then, we fit a two-dimensional second-order spline to each integration on a per-amplifier basis to the unmasked portions of each image. We extrapolated the combined background spline for the whole image over the masked portions near WASP-80 and subtracted it from the original image values. In roughly 3\% of the integrations, the reference pixel correction failed for at least one of the amplifier regions, so after the spline fitting and subtraction, we re-ran the reference pixel correction using \texttt{hxrg-ref-pixel}\footnote{\url{https://github.com/JarronL/hxrg_ref_pixels}}. Finally, we attempted to remove some of the red-noise caused by the NIRCam readout electronics, which is present along detector rows, by calculating the robust mean of each row using pixels from column 1800 onwards for the F322W2 images and up to column 600 for F444W (both chosen to avoid light from WASP-80), and then subtracting this mean from each row.

Next, we measured spectroscopic light curves from our background-subtracted images, using optimal extraction to measure the 1D spectrum in each image. We extracted spectroscopic lightcurves from 2.45\,$\mu$m to 3.95\,$\mu$m at F322W2 and from 3.89\,$\mu$m to 4.97\,$\mu$m at F444W, both using 0.015\,$\mu$m-wide spectral channels. We linearly interpolated over each spectral column during this extraction process to account for partial-pixel effects in the 0.015\,$\mu$m wavelength bins.

We fit each spectroscopic lightcurve using a \texttt{BATMAN} eclipse model \cite{kreidberg2015batman}. We did not discard any of the data points at the beginning of the lightcurves. We fixed the orbital parameters of WASP-80 b to those measured in the \texttt{Eureka!} joint broadband fit tabulated in Table \ref{tab:orbitalParameters}, which left the free parameters in our spectroscopic lightcurve fitting to be the secondary eclipse depth and the slope and normalization of a background linear trend. We did not impose a prior on any of these parameters. We fit each spectral channel individually.

We performed an initial likelihood maximization using a Nelder-Mead sampler followed by Markov chain Monte Carlo (MCMC) likelihood sampling to fit the spectroscopic lightcurves. We used the maximum likelihood point identified by the Nelder-Mead maximization as the starting locus for initializing the MCMC chains. To perform the MCMC runs, we used the \texttt{emcee} Python package \citep{foreman-mackey2013emcee} using 12 walkers with a 2,000-step burn-in and then a 4,000-step production run for each spectral channel. We checked that the MCMC had converged by verifying that the Gelman-Rubin statistic \cite{GelmanRubin1992} was below 1.1 for each parameter in each spectral channel.

We additionally checked the goodness-of-fit and statistical properties of our eclipse modeling in each spectral channel. We verified that the average of the per-point flux uncertainties in each channel's lightcurve matched the standard deviation of the residuals to the best-fit eclipse model. We also computed the Anderson-Darling statistic for each channel's lightcurve residuals to check that the residuals appeared Gaussian. We did not find statistically significant non-Gaussianity in the residuals of our spectroscopic lightcurve fits.

\subsection*{Modeling Approach}
Inferring the properties (composition, cloud properties, and thermal structure) of exoplanet atmospheres from spectroscopic observations necessitates data-model comparisons via statistical methods, commonly referred to as ``retrievals." In this work, we utilize both ``grid-based" and ``free" retrievals. Our grid-based retrievals start with a set of pre-computed atmosphere models in radiative-convective-photochemical equilibrium (RCPE) over a defined parameter space from which we then estimate values for each parameter that best agrees with the spectral observations. In contrast, free retrievals are not burdened by restrictive RCPE assumptions; instead, they only estimate parameter values that achieve the closest data-model fit. Our consideration of both free and grid-based retrievals enables a comprehensive analysis -- grid-based retrievals with RCPE models force physical solutions, while free retrievals can illuminate where physics in the grid models does not sufficiently capture spectral features and provide possible explanations. We use the CHIMERA modeling framework \cite[initially presented in ][]{Line2013} to produce atmosphere models and their resulting eclipse spectra, and we use Bayesian nested sampling to infer planet properties. We follow a similar modeling process to those presented in refs. \cite{Bell2023, Welbanks2024, Beatty2024}.  

\begin{table}
    \centering
    \begin{tabular}{c|c}
        \hline\hline Parameter & Grid Values \\
        \hline T$_{\rm day}$ (K) & 825, 850, 875, 900 \\ 
        T$_{\rm int}$ (K) & 150, 200, 250, 300, 350, 400, 450 \\ 
        $[$M/H$]$ & 0.375, 0.5, 0.625, 0.75, 0.875, 1.0, 1.125,\\
         & 1.25, 1.375 \\
        C/O &  0.3, 0.4, 0.5, 0.6, 0.7 \\ 
        log$_{10}$(K$_{\rm zz}$) (cm$^2$s$^{-1}$)& 8, 8.5, 9, 9.5, 10, 10.5, 11, 11.5 \\ \hline \hline
    \end{tabular}
    \caption{The parameter space for ScCHIMERA grid models.}
    \label{tab:grid}
\end{table}

\subsection*{Grid-Based Retrieval}
Using the Self-consistent \texttt{CHIMERA} code, \texttt{ScCHIMERA} (recently presented in detail in refs. \cite{Iyer2023, Wiser2024}), coupled with the kinetics code, \texttt{VULCAN} \cite{Tsai2017}, we generate a grid of 1D atmosphere models in radiative-convective-photochemical equilibrium (1D-RCPE). The following parameters are varied to produce a model grid: planet dayside temperature via scaled incident stellar flux (T$_{\rm day}$), internal temperature (T$_{\rm int}$), chemical composition via metallicity and the carbon-to-oxygen ratio, and vertically constant mixing strength in the atmosphere via the eddy diffusion parameter K$_{\rm zz}$. See Table \ref{tab:grid} for a list of the grid parameter space. 

To produce an individual model, \texttt{ScCHIMERA} first iteratively solves for the radiative-convective equilibrium pressure-temperature profile from an internal temperature and the top-of-atmosphere incident stellar flux, which we compute from a \texttt{PHOENIX} stellar model (T$_{star}$~=~4143 K, log$_{10}$(g)[c.g.s]~=~4.663) \cite{Husser2013}. From the metallicity ([M/H]) and the carbon-to-oxygen ratio (C/O), \texttt{ScCHIMERA} computes the thermochemical equilibrium gas volume mixing ratios using the \texttt{NASA CEA2} routine for Gibbs free energy minimization \cite{GordonMcbride1994} along the pressure-temperature profile, producing an atmosphere model in radiative-convective-thermochemical equilibrium (1D-RCTE). This equilibrium chemistry routine solves for thousands of molecular/atomic species; however, we only include opacity sources for major radiative species (H$_2$/He collision-induced absorption, H/e-/H- bound/free-free continuum, and the line opacities for H$_2$O, CO, CO$_2$, CH$_4$, NH$_3$, H$_2$S, PH$_3$, HCN, C$_2$H$_2$, OH, TiO, VO, SiO, FeH, CaH, MgH, CrH, ALH, Na, K, Fe, Mg, Ca, C, Si, Ti, O, Fe+, Mg+, Ti+, Ca+, C+).

The pressure-temperature profile and associated gas volume mixing ratios (CH$_4$, CO, CO$_2$, C$_2$H$_2$, H, HCN, He, H$_2$, H$_2$O, H$_2$S, NH$_3$, and N$_2$) for each 1D-RCTE model are then passed into \texttt{VULCAN} to solve for radiative-convective-photochemical equilibrium (1D-RCPE), the chemical state of the atmosphere once accounting for photochemistry and vertical mixing. We use the H-C-O-N-S kinetics network presented in ref. \citep{Tsai2023}, a vertically constant eddy diffusion (K$_{\rm zz}$) profile, and the UV-stellar spectrum from GJ676A (an M0V star, part of the Mega-MUSCLES survey \cite{Wilson2021}) as a proxy for WASP-80. \texttt{VULCAN} iteratively converges on modified mixing ratios, which we then input back into \texttt{ScCHIMERA}, fixing their abundances and again computing 1D-RCTE for all other species not included in \texttt{VULCAN}. We follow this \texttt{ScCHIMERA}-to-\texttt{VULCAN}-to-\texttt{ScCHIMERA} process again to reach convergence, i.e., so that the pressure-temperature profile and gas volume mixing ratios are not significantly changing. This results in a grid of 1D-RCPE models that were modified in \texttt{VULCAN} two times.\footnote{This atmosphere model grid is available on Zenodo: https://zenodo.org/records/14884687} 

We estimate WASP-80 b's atmosphere parameters using nested sampling with \texttt{PyMultiNest} \cite{Buchner2014} and 500 live points (parameter combinations) over the full model grid (T$_{\rm day}$, T$_{\rm int}$, [M/H], C/O, and log$_{10}$(K$_{\rm zz}$)), with the addition of a vertically uniform grey cloud opacity ($\kappa _{\rm cld}$, effectively an abundance weighted grey cross-section), and a dilution factor ($A$) multiplying the planetary flux spectrum to account for temperature heterogeneities across the dayside disk \cite[e.g.,][]{Taylor2020}. We penta-linearly interpolate (using \texttt{SciPy RegularGridInterpolator}) the pressure-temperature profile and gas volume mixing ratios for a given set of parameters, then post-process models into cloudy emission spectra within the \texttt{PyMultiNest} routine at a spectral resolution of 100,000 (including opacities for H$_2$-H$_2$/He CIA, H$_2$O, CO, CO$_2$, CH$_4$, NH$_3$, HCN, C$_2$H$_2$, H$_2$S, SO$_2$, Na, and K) to avoid any resolution-linked biases. 

\begin{figure}[t]
\centering
\includegraphics[width=1.0\linewidth]{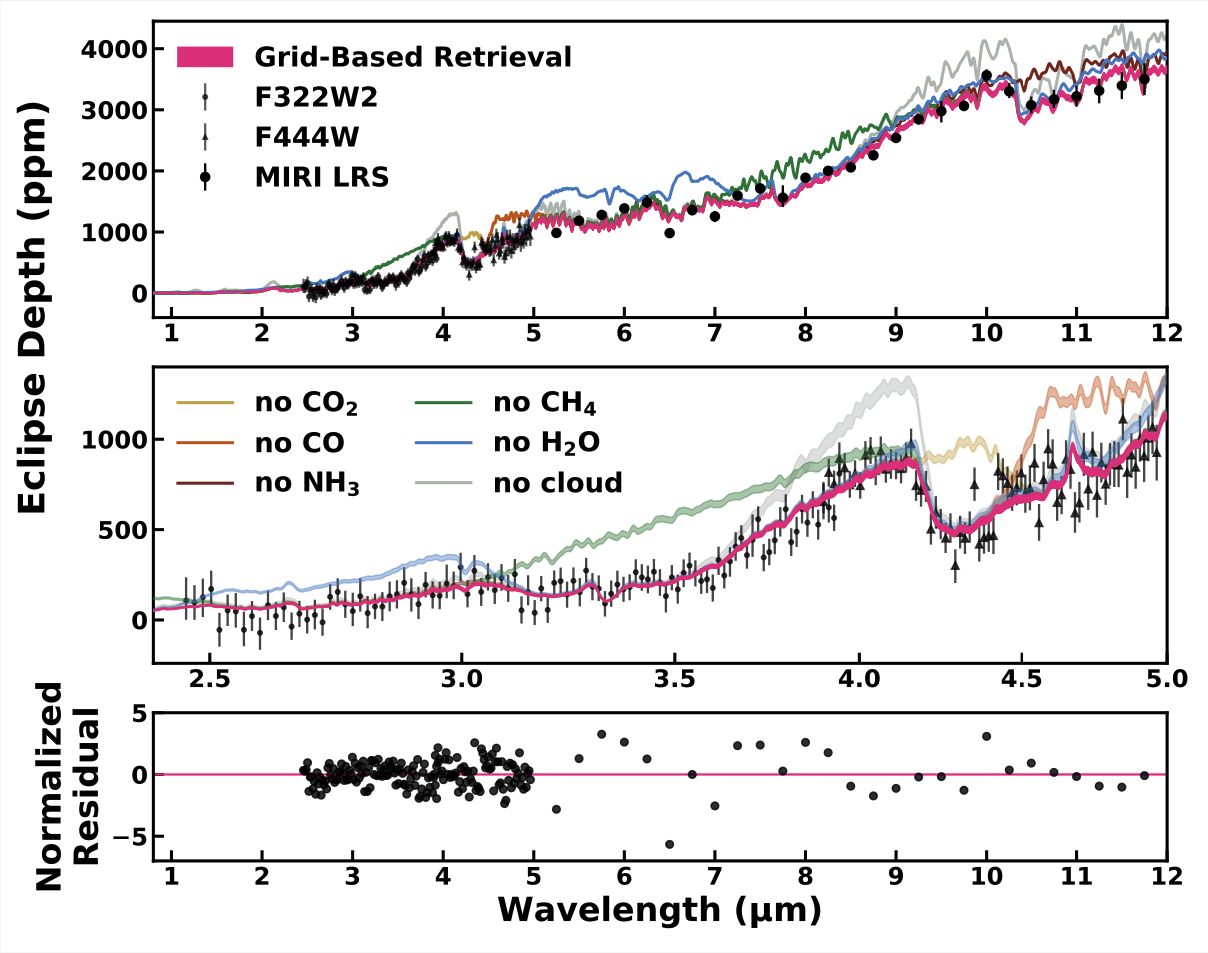}
\caption{\textit{Top and Middle:} Modeled emission spectra from the grid-based retrieval. The middle plot zooms in on only the NIRCam observations. \textbf{Pink:} The 2$\sigma$ confidence region of the emission spectrum from the grid-based retrieval. \textbf{Other colored lines:} The grid-based retrieval spectrum with individual molecules, or the uniform grey cloud, turned “off,” illustrating their spectral signatures. On the top, only the median spectrum for each removed molecule/cloud is shown, while in the middle, the 2$\sigma$ confidence region is plotted. \textit{Bottom:} Normalized residuals ([data-model]/error) comparing the best-fit grid spectrum to the observations.}
\label{fig:spec}
\end{figure}

\subsection*{Free Retrieval}
To complement parameter estimates from the RCPE grid-based retrievals, we also perform more flexible ``free" retrievals with \texttt{CHIMERA}. In this process, a model spectrum is generated given individual constant-with-altitude molecular gas mixing ratios, a parametric vertical temperature profile, and cloud properties without forcing RCPE assumptions.

We infer planet parameters with \texttt{PyMultiNest} nested sampling \cite{Buchner2014} and 500 live points. \texttt{CHIMERA} models a one-dimensional hydrostatic equilibrium atmosphere spanning pressures 10$^{-6}$ to 10$^{1.2}$ bar in 10$^{0.1}$ bar layers. The vertical temperature structure is parameterized following the prescription from \cite{MadhuSeager2009}, and independent free parameters are included for gas volume mixing ratios of species expected in warm exoplanet atmospheres: H$_2$O, CO, CO$_2$, CH$_4$, NH$_3$, and SO$_2$ \cite{Tsai2022, Schlawin2024}. The model performs line-by-line opacity sampling at a spectral resolution of 100,000 before being binned to the resolution of the observations. Finally, we include $\kappa _{\rm cld}$ and $A$, as we do in the model grid, allowing for a vertically uniform grey cloud opacity. Overall, we include 14 free parameters: 6 molecules, 6 pressure-temperature variables, $\kappa _{\rm cld}$, and $A$.

\begin{SCfigure*}[\sidecaptionrelwidth][t]
\centering
\includegraphics[width=11.4cm]{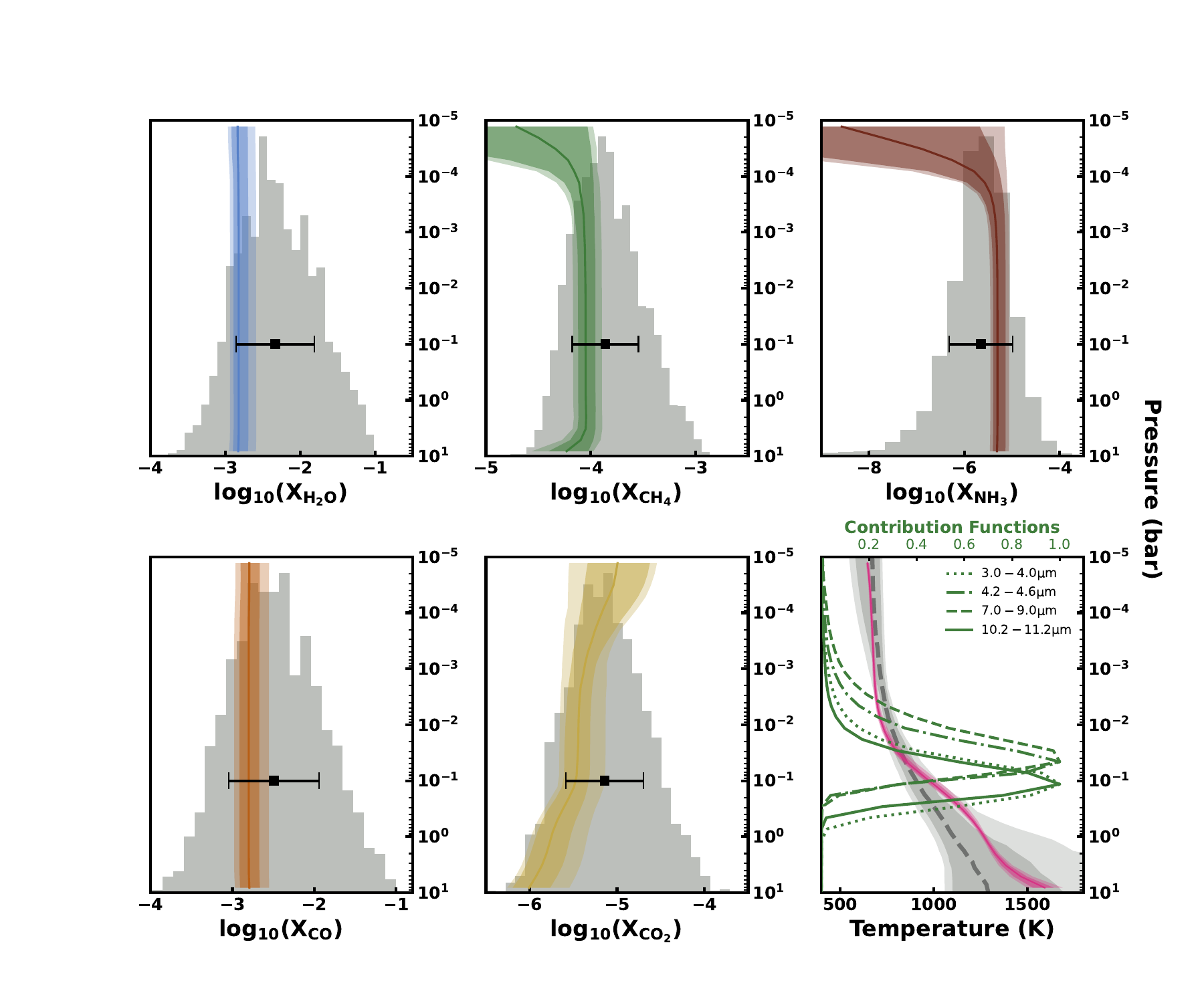}
\caption{Molecular abundance estimates and the pressure-temperature profile from the free (grey posteriors) and grid-based retrievals (colored abundance profiles with pressure). Black points indicate free retrieval posterior medians and 1$\sigma$ confidence regions. CO$_2$, CO, CH$_4$, and H$_2$O are confidently detected to $>$~7.5$\sigma$. NH$_3$ is non-decisively detected at 2.8$\sigma$. SO$_2$ is not plotted because the abundance is log$_{10}$($X_{\rm SO_2}$)~$<$~-6.77 in both the free and grid-based retrievals. All plotted molecular abundances are consistent with those in \cite{Bell2023} with narrower posterior distributions. Contribution functions highlighting the pressures probed by our observations are shown in green on the pressure-temperature panel.}
\label{fig:abund}
\end{SCfigure*}

\section*{Results}
Parameter estimations from the grid-based and free retrievals are presented in Table \ref{tab:estimates}, and posterior distributions for each retrieval can be found in Supporting Figures S5--S6. To evaluate the goodness-of-fit of our best-fit spectra from each retrieval, we calculated p-values for $\chi^2$ and the Kolmogorov-Smirnov (K-S) test, which assesses whether the normalized residuals ([data-model]/error) follow a Gaussian distribution, i.e., that the residuals may result from random measurement error. The grid-based retrieval returns $\chi^2/N_{\rm DOF}=$ 1.38 with p-value $=4\times10^{-4}$; however, the K-S p-value~=~0.89 indicates high confidence that the residuals are drawn from a normal distribution. Similarly, the free retrieval returns $\chi^2/N_{\rm DOF}=$~1.36, $\chi^2$ p-value~$=8\times10^{-4}$, and K-S p-value~=~0.86. Notably, for the grid-based retrieved spectrum, $\chi^2/N_{\rm DOF}=$~0.95 when only considering the NIRCam data and $\chi^2/N_{\rm DOF}=$~5.38 for the MIRI LRS observations. To test the reliance of our parameter inferences on the MIRI LRS observations, we performed a grid-based retrieval to only the NIRCam observations and found consistent results. Spectra and the posterior distribution for this retrieval can be found in Supporting Information Figures S8--S9, and a complete table of fit statistics can be found in Supporting Information Table S2.

\begin{table}[ht!]
    \centering
    \begin{tabular}{c|c|c}
        \hline\hline Parameter & Estimate & Prior (low, high)\\
        \hline \multicolumn{3}{c}{Grid-Based Retrieval} \\
        \hline T$_{\rm day}$ (K) & 859.87$^{+7.08}_{-9.10}$ & (825, 900)\\ 
        T$_{\rm int}$ (K) & 381.08$^{+37.70}_{-38.98}$ & (150, 450) \\ 
        $[$M/H$]$ & 0.55$^{+0.12}_{-0.10}$ & (0.375, 1.375)\\
        C/O & 0.48$^{+0.06}_{-0.07}$ & (0.3, 0.7)\\ 
        log$_{10}$(K$_{\rm zz}$) (cm$^2$s$^{-1}$) & 9.13$^{+1.06}_{-0.74}$ & (8.0, 11.5)\\ 
         & 2$\sigma$ upper limit at 11.07 & \\
        log$_{10}$($\kappa_{\rm cld}$) & -29.50$^{+0.06}_{-0.05}$ & (-35, -20) \\ 
        dilution factor ($A$) & 1.14$^{+0.03}_{-0.03}$ & (0.5, 1.5) \\ \hline 
        \multicolumn{3}{c}{Free Retrieval} \\
        \hline log$_{10}$($X_{\rm H_{2}O}$) & -2.33$^{+0.59}_{-0.52}$ & (-12, -1)\\ 
        log$_{10}$($X_{\rm CO}$) & -2.49$^{+0.60}_{-0.55}$ & (-12, -1) \\ 
        log$_{10}$($X_{\rm CO_{2}}$) & -5.14$^{+0.49}_{-0.45}$ & (-12, -1) \\
        log$_{10}$($X_{\rm CH_{4}}$) & -3.86$^{+0.37}_{-0.31}$ & (-12, -1) \\ 
        log$_{10}$($X_{\rm NH_{3}}$) & -5.65$^{+0.57}_{-0.67}$ & (-12, -1) \\ 
        log$_{10}$($X_{\rm SO_{2}}$) & -9.47$^{+1.66}_{-1.61}$ & (-12, -1) \\ 
        & 2$\sigma$ upper limit at -6.77 & \\ 
        T$_{\rm 1\mu bar}$(K) & 667.79$^{+43.19}_{-103.96}$ & (300, 1200) \\ 
        $\alpha_{1}$ (K$^{-1/2}$) & 1.07$^{+0.61}_{-0.45}$ & (0.02, 2.0) \\ 
        $\alpha_{2}$ (K$^{-1/2}$) & 0.40$^{+0.17}_{-0.13}$ & (0.02, 2.0)\\ 
        log$_{10}$P$_1$ (bar) & -1.75$^{+1.72}_{-0.67}$ & (-6, 2) \\
        log$_{10}$P$_2$ (bar) & -3.89$^{+1.32}_{-1.24}$ & (-6, 2) \\
        log$_{10}$P$_3$ (bar) & 0.95$^{+0.67}_{-0.84}$ & (-6, 2) \\
         & 2$\sigma$ lower limit at -0.49 &  \\
        log$_{10}$($\kappa_{\rm cld}$) & -32.74$^{+1.53}_{-1.46}$ & (-35, -20) \\ 
         & 2$\sigma$ upper limit at -30.22 & \\ 
        dilution factor ($A$) & 1.09$^{+0.04}_{-0.04}$ & (0.5, 1.5) \\ \hline \hline
    \end{tabular}
    \caption{Grid-based and free retrieval parameter estimates and priors. Parameters are: dayside temperature (T$_{\rm day}$), internal temperature (T$_{\rm int}$), log$_{10}$ metallicity relative to solar abundances ($[$M/H$]$), carbon-to-oxygen ratio (C/O), eddy diffusion strength (log$_{10}$(K$_{\rm zz}$)), grey cloud opacity (log$_{10}$($\kappa_{\rm cld}$)), dilution factor ($A$), molecular volume mixing ratios (log$_{10}$($X_{\rm x}$)), and pressure-temperature profile parameters \cite[T$_{1\mu bar}$, $\alpha_{1}$, $\alpha_{2}$, log$_{10}$P$_1$, log$_{10}$P$_2$, log$_{10}$P$_3$, ][]{MadhuSeager2009}.}
    \label{tab:estimates}
\end{table}

Both the free and grid-based retrievals strongly detect H$_2$O, CO, CO$_2$, and CH$_4$ (see Figure \ref{fig:spec} for the retrieved grid-based spectrum and Supporting Information Figure S7 for the free retrieval spectrum). From a Bayesian evidence comparison of free retrievals with and without each molecule present, we confidently detect all four molecules at greater than 7.5$\sigma$: H$_2$O at 13.0$\sigma$, CO at 7.5$\sigma$, CO$_2$ at 10.0$\sigma$, and CH$_4$ at 15.1$\sigma$. The free retrieval additionally detects NH$_3$ at 2.8$\sigma$, which we consider a possible but non-conclusive detection. SO$_2$ is not detected with only a 2$\sigma$ upper limit at log$_{10}$($X_{\rm SO_2}$)~=~-6.92. We also compute detection significances from the grid-based retrieval by comparing our preferred model to retrievals with one molecule ``turned off" in the grid spectra. The grid-based retrieval returns higher detection confidence than the free retrieval for all molecules; however, it is important to note that grid models with individual molecules turned off are no longer in RCPE. Additionally, free retrievals may compensate for the lack of one molecule with other parameters, but this is not possible for grid-based retrievals, so the molecule detection significances computed from the free retrievals are more conservative. Gas volume mixing ratio constraints for H$_2$O, CO, CO$_2$, CH$_4$, NH$_3$, and SO$_2$ from the free retrievals are consistent with abundances in the grid models (see Figure \ref{fig:abund}). 

The grid-based retrieval prefers the inclusion of a vertically uniform grey cloud opacity, log$_{10}$($\kappa_{\rm cld}$)~=~-29.50$^{+0.06}_{-0.05}$, while the free retrieval predicts a 2$\sigma$ upper limit on clouds at log$_{10}$($\kappa_{\rm cld}$)~=~-30.22, also allowing for a cloud-free atmosphere (see Supporting Information Figures S6--S7). As a result, the presence of clouds is inconclusive. However, clouds are not unreasonable for $T_{\rm day}=$~895.25$^{+3.16}_{-3.19}$ K \cite[e.g., ][]{Gao2021}. 

The grid-based retrieval also predicts vertical mixing (2$\sigma$ upper limit at $\log_{10}$(K$_{\rm zz}$)~=~11.07~cm$^2$s$^{-1}$), along with a high internal temperature, T$_{\rm int}$~=~381$^{+38}_{-39}$~K. This combination serves to quench CH$_4$. Recent JWST transmission analyses of WASP-107 b, a 750 K Neptune-sized planet, report a similar effect, with ref. \cite{Welbanks2024} estimating $\log_{10}$(K$_{\rm zz}$)=8.4-9.0~cm$^2$s$^{-1}$ and T$_{\rm int}$~$>$~345~K, and ref. \cite{Sing2024} estimating $\log_{10}$(K$_{\rm zz}$)~=~11.6$\pm$0.1~cm$^2$s$^{-1}$ with T$_{\rm int}$~$=$~460$\pm$40~K to explain depleted CH$_4$ abundances. Using general circulation models, ref. \cite{Komacek2019} estimates that $\log_{10}$(K$_{\rm zz}$)~$\approx$~8.0-10.0 cm$^2$s$^{-1}$ may be expected for gas giants in the temperature regime of WASP-107 b and WASP-80 b. To test the necessity of including free parameters for internal temperature and eddy diffusion strength in our analysis of WASP-80 b, we conducted a grid-based retrieval fixing T$_{\rm int}$~=~150~K and $\log_{10}$(K$_{\rm zz}$)~=~9.0~cm$^2$s$^{-1}$. These assumptions result in a low C/O estimate (2$\sigma$ upper limit at C/O~=~0.31) that aims to reduce the CH$_4$ abundance; however, low C/O also limits CO$_2$ and the resulting modeled spectrum does not fit the CO$_2$ feature centered at 4.3$\mu$m. The model that includes free parameters for T$_{\rm int}$ and $\log_{10}$(K$_{\rm zz}$) is preferred by 7.3$\sigma$.

\section*{Discussion}

\begin{figure}[t]
    \centering
    \includegraphics[width=1.\linewidth]{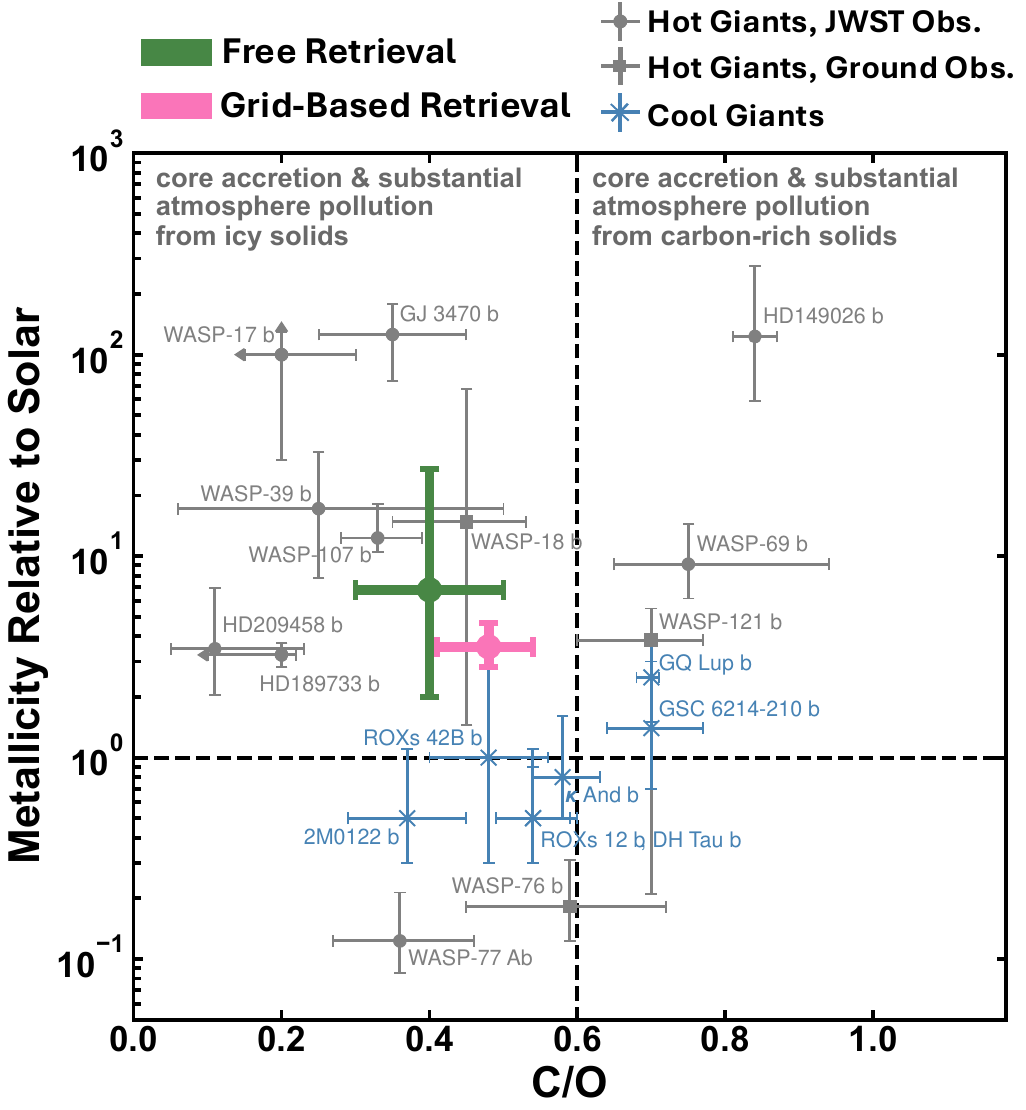}
    \caption{Metallicity and C/O estimates for WASP-80 b and other gas giant planets and brown dwarfs. Planets shown in grey are close-in transiting gas giants. Composition inferences shown with circular points are drawn from JWST observations, while square points indicate ground-based high-resolution observations. Blue points are cool gas giants and brown dwarfs ($\sim$10-30 M$_{\rm Jupiter}$) with wide orbits \cite{Xuan2024}. Dotted lines denote approximate solar abundances. WASP-80 b's [M/H] and C/O are consistent with other hot gas giants thought to have formed via core accretion and disk migration through an undissipated disk. The plotted wide orbit companions, in contrast, have metallicities more consistent with solar elemental abundances. Metallicity and C/O estimates come from refs. \cite{Xuan2024, Fu2024, Xue2024, Brogi2023, Bean2023, August2023, WeinerMansfield2024, Welbanks2024, Schlawin2024, Beatty2024, Gressier2024, Smith2024} and Welbanks et al. in prep.}
    \label{fig:formation}
\end{figure}

WASP-80 b's metallicity and C/O estimates are consistent with other hot gas giant exoplanets. [M/H] and C/O are free parameters in the grid-based retrieval, and they can be estimated from molecular abundance constraints in the free retrieval. In Figure \ref{fig:formation}, we plot WASP-80 b's metallicity and C/O estimates from this work compared to other gas giants, both hot transiting planets (grey points) and cooler giants planets and brown dwarf companions (blue points). We also mark approximate solar abundance ratios. The host star, WASP-80, has a metallicity consistent with solar to slightly sub-solar metallicity ([Fe/H]~=~$-0.13^{+0.15}_{-0.17}$ \cite{Triaud2015}).

WASP-80 b's super-solar metallicity is consistent with other hot gas giant planets that are hypothesized to have formed via core accretion beyond the H$_2$O ice line, followed by the growth of an enhanced metallicity atmosphere through the accretion of H-depleted gas and metal-rich ices during disk migration inwards to a close-in orbit. WASP-80 b's solar to sub-solar C/O indicates that there has not been substantial pollution of the atmosphere from carbon-rich grains, but rather a combination of these grains and oxygen-rich ices. 

In contrast, gas giants and brown dwarfs ($\sim$10--30 M$_{\rm Jupiter}$) with wide orbital separations have metallicities closer to solar abundances \cite{Xuan2024}. This is consistent with formation beyond ice lines, either via direct gravitational collapse \cite{Boss1997, Durisen2007, Burn2021, Xuan2024} or core accretion beyond the CO ice line \cite[e.g.,][]{Chachan2023}. The lack of migration prevents substantial metal enrichment via polluting material. Ref. \cite{Wang2023}, however, shows that less massive gas giants ($\sim$2--10 M$_{\rm Jupiter}$) at wide orbital separations may have super-solar metallicities ($\sim$0.1--0.7 [M/H]), not dissimilar from some hot gas giants, without migration inwards. The complexity of planet formation may mean that mapping metallicity and C/O measurements neatly back to a formation pathway is fraught \cite{Welbanks2019, Chachan2023}, but existing population trends show that WASP-80 b's composition is consistent with other hot gas giants.

Giant planets, like WASP-80 b, are rare around low-mass stars. This is likely because protoplanetary disks around low-mass stars do not contain enough mass to easily facilitate their formation via core accretion. Ref.\ \cite{Burn2021} provides a possible formation pathway for low stellar mass Jovians through a combination of protoplanetary disk mass on the high end of what is expected (from ref.\ \cite{Burn2021}, this may be on the order of M$_{\rm gas}$ $>$ 0.007 M$_{\rm Sun}$ and M$_{\rm solid}$ $>$ 66 M$_{\rm Earth}$) and slow type I planet migration, i.e., migration that occurs before the planet is massive enough to open a gap in the protoplanetary disk \cite{Alibert2005, Mordasini2009a}. Ref.\ \cite{Burn2021} additionally show that massive planets around low-mass stars may be more likely to arise if their planetesimals form nearer to the star due to higher disk mass densities closer to the star. Ref.\ \cite{SavBit2024} suggests that disk-mass problems may be explained by underestimated disk-mass estimates due to inaccurate assumptions in modeling disk observations, including the use of a flux-to-mass conversion. Additionally, planet formation beginning earlier (before a well-defined disk, i.e., pre-Class II disk) may mean more available mass for gas giant formation \cite{SavBit2023}. 

\subsection*{Exoplanets with JWST}
Studying the atmospheres of exoplanets with transit spectroscopy is one of the primary methods by which we build our understanding of their compositions, climates, and formation pathways, providing context for our own unique solar system. In the past $\sim$25 years of exoplanet atmosphere observations, most of what we have learned has come from the Spitzer and Hubble Space Telescopes. In its first few years of operation, JWST is already enabling new insights into exoplanet atmospheres beyond those achieved with Spitzer and Hubble. About 100 exoplanet systems have already been observed with JWST, or are planned to be, with increased wavelength coverage and spectral resolution compared to previous space-based observatories. These observations are enabling detections of new exoatmospheric gases (e.g., \cite{Bell2024}) and evidence of climate processes, such as 3D heat circulation (e.g., \cite{Coulombe2023ERS}) and tidal heating (e.g., \cite{Welbanks2024}). 

Observing planets with atypical characteristics provides a valuable opportunity to stress test our assumptions and refine our predictions about planet formation processes. In this paper, we present a panchromatic emission spectrum for WASP-80 b, the first gas giant around a late-K/early M-dwarf and the coolest planet for which JWST has obtained a complete emission spectrum 2.4--10 $\mu$m. We constrain the abundances of five major chemical species in the atmosphere, significantly narrowing abundance constraints previously published from NIRCam F332W2 observations alone \cite{Bell2023}, and connect those constraints to the planet's metallicity, C/O, and formation. In the coming years, we expect that transit spectroscopy with JWST will continue to enable new insights into atmospheres and formation pathways for diverse populations of exoplanets.

\acknow{%
We acknowledge Research Computing at Arizona State University for providing high-performance computing and storage resources that have significantly contributed to the research results reported within this manuscript. L.S.W. and M.R.L. acknowledge support from NASA/STScI award HST-AR-17050 and JWST-ERS-01366. T.J.B. and T.P.G. acknowledge funding support from the NASA Next Generation Space Telescope Flight Investigations program (now JWST) via WBS 411672.07.04.01.02 and 411672.07.05.05.03.02. We thank Marcia Rieke (NIRCam PI) for providing the NIRCam observing time. This work benefited from the 2024 Exoplanet Summer Program in the Other Worlds Laboratory (OWL) at the University of California, Santa Cruz, a program funded by the Heising-Simons Foundation and NASA.}

\showacknow{} 

\bibsplit[12] 


\bibliography{main}

\SetWatermarkText{}  


\section*{Supplementary material}

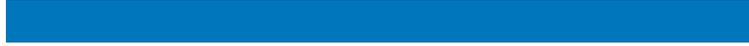

# Supporting Information for

A Precise Metallicity and Carbon-to-Oxygen Ratio for a Warm Giant Exoplanet from its Panchromatic JWST Emission Spectrum

Lindsey S. Wiser, Taylor J. Bell, Michael R. Line, Everett Schlawin, Thomas G. Beatty, Luis Welbanks, Thomas P. Greene, Vivien Parmentier, Matthew M. Murphy, Jonathan J. Fortney, Kenny Arnold, Nishil Mehta, Kazumasa Ohno, and Sagnick Mukherjee

Lindsey S. Wiser
E-mail: lindsey.wiser@asu.edu

**This PDF file includes:**

Supporting text
Figs. S1 to S9
Tables S1 to S2
SI References

**Supporting Information Text**

**Eclipse Timing Posteriors.** We ran an additional fit to confirm the consistency of the three eclipse timings with a $e = 0$, phase=0.5 prediction. We performed a joint fit of the three eclipse observations (each with an independently fit mid-eclipse time), the published NIRCam/F322W2 transit (from ref. (1)), and a Gaussian prior on the orbital parameter posteriors of ref. (2). The same semi-major axis ($a$), orbital inclination ($i$), and orbital period ($P$) parameters were used for all observations. Only the ref. (2) posterior and the NIRCam/F322W2 transit observation were used to constrain the mid-transit time parameter ($t_0$) from which we computed the $e = 0$, phase=0.5 timing using $t_{\rm ecl} = t_0 + P/2 + ltt$. We computed the difference in light travel time between mid-transit and mid-eclipse ($ltt$) as $34.318 \pm 0.095$ seconds. The per-instrument eclipse timings were constrained only by that instrument's observations and a weakly informative prior of $2{,}456{,}488.959 \pm 0.010$ BJD$_{\rm TDB}$. Our findings are summarized in Table S1 and Figure S4, which confirm that the assumption of zero eccentricity is statistically consistent with the precision of the observations, and that the mid-eclipse times from each epoch are statistically consistent with each other.

Lindsey S. Wiser, Taylor J. Bell, Michael R. Line, Everett Schlawin, Thomas G. Beatty, Luis Welbanks, Thomas P. Greene, Vivien Parmentier, Matthew M. Murphy, Jonathan J. Fortney, Kenny Arnold, Nishil Mehta, Kazumasa Ohno, and Sagnick Mukherjee

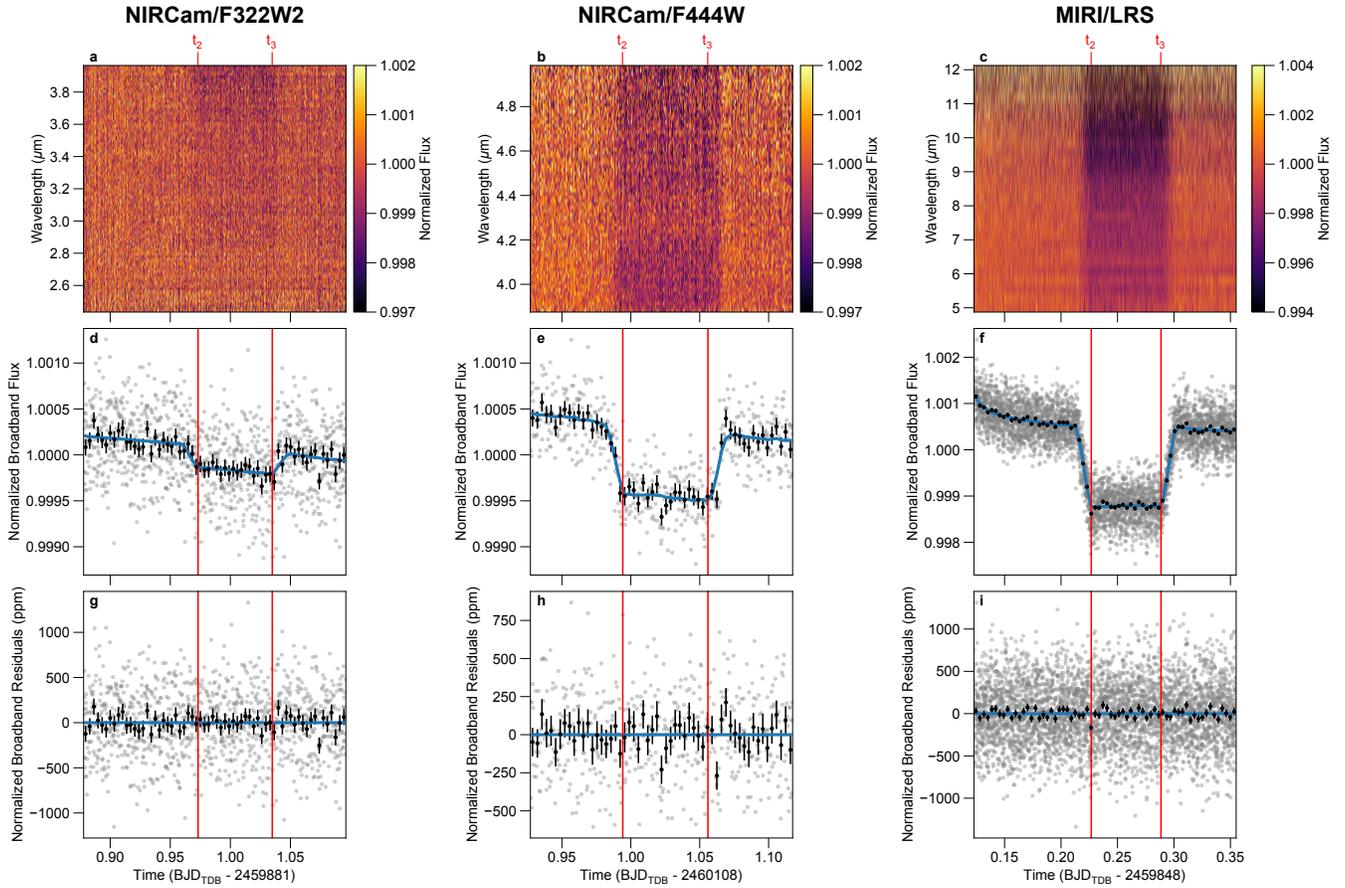

**Fig. S1.** The normalized spectroscopic lightcurves (panels a–c) and broadband lightcurves (panels d–f) from our fiducial `Eureka!` reduction. In panels d–f, the grey points show raw broadband flux measurements, black points with error bars show ∼5 minute binned data (for illustration purposes only), blue line shows the fitted broadband model, and red vertical lines show the second and third contacts of the eclipses. Panels g–i show the residuals of the fit to the broadband data.

Lindsey S. Wiser, Taylor J. Bell, Michael R. Line, Everett Schlawin, Thomas G. Beatty, Luis Welbanks, Thomas P. Greene, Vivien Parmentier, Matthew M. Murphy, Jonathan J. Fortney, Kenny Arnold, Nishil Mehta, Kazumasa Ohno, and Sagnick Mukherjee

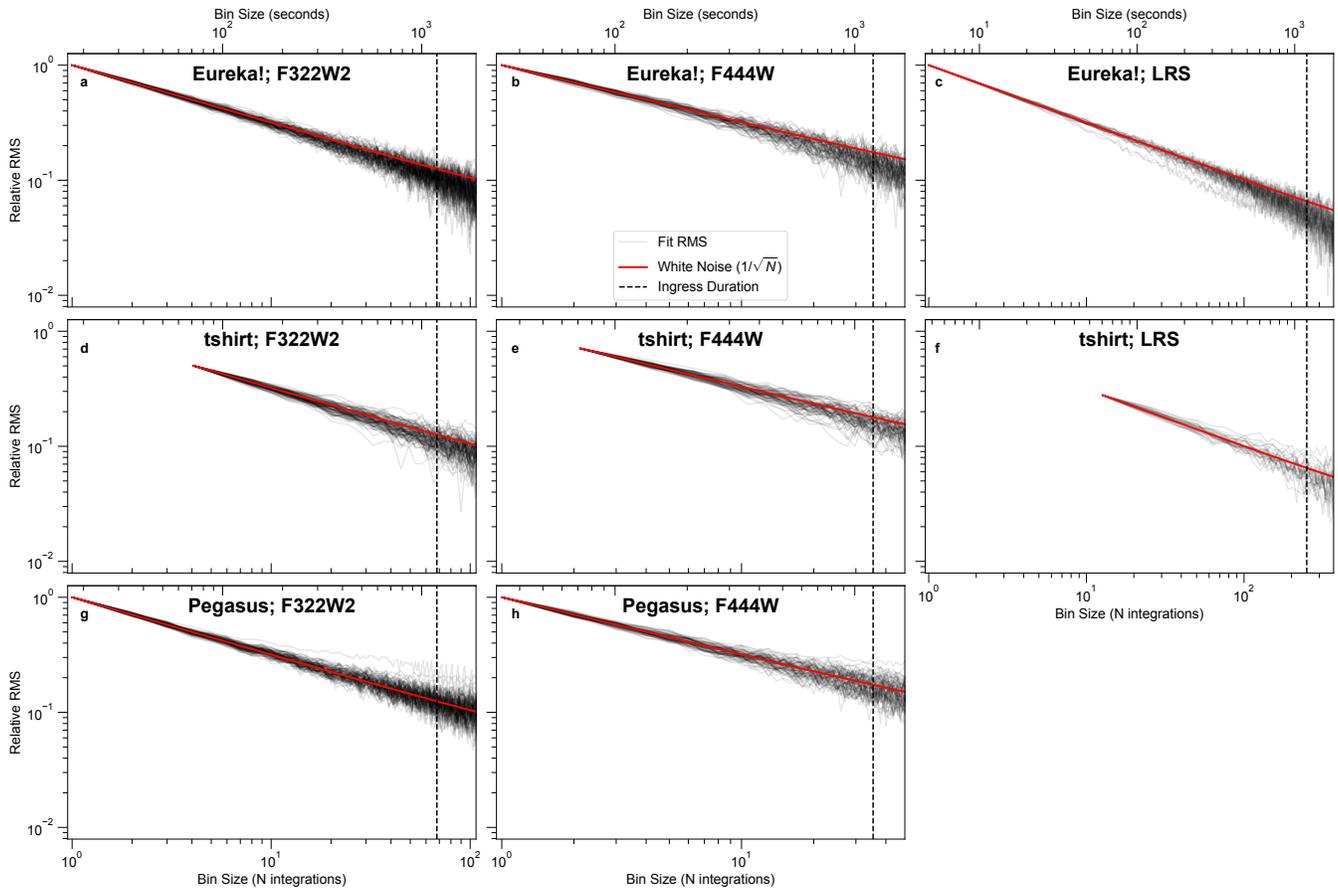

**Fig. S2.** Allan variance plots (3) for the spectroscopic fits from all three instruments and all three reduction pipelines. With the exception of a small number of channels in the `Pegasus` NIRCam/F322W2 reduction and the `tshirt` MIRI/LRS reduction, there appears to be little-to-no residual red noise. Note that the Allan variance plot for `Eureka!`'s LRS reduction has had the GP model removed from the residuals, without which the Allan variance plot for some LRS spectral channels exhibited residual red noise.

Lindsey S. Wiser, Taylor J. Bell, Michael R. Line, Everett Schlawin, Thomas G. Beatty, Luis Welbanks, Thomas P. Greene, Vivien Parmentier, Matthew M. Murphy, Jonathan J. Fortney, Kenny Arnold, Nishil Mehta, Kazumasa Ohno, and Sagnick Mukherjee

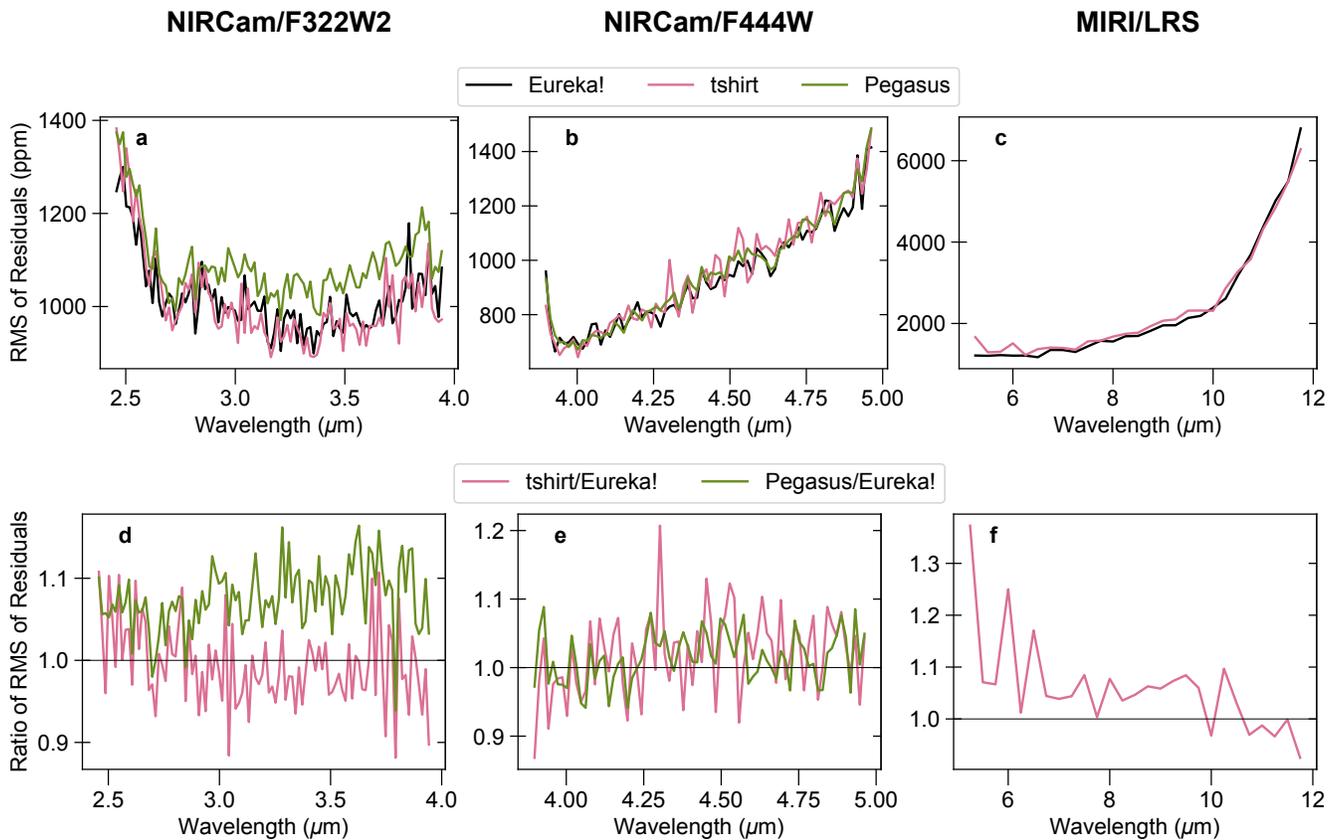

**Fig. S3.** *Top:* A comparison of the root-mean-square (RMS) of the residuals from the spectroscopic fits for all three instruments (*a*: NIRCam/F322W2, *b*: NIRCam/F444W, and *c*: MIRI/LRS) and all three reduction pipelines (black: `Eureka!`, pink: `tshirt`, and green: `Pegasus`). *Bottom*: The root-mean-square (RMS) of the spectroscopic residuals from the `tshirt` and `Pegasus` reductions compared to the `Eureka!` reduction for all three instruments (*d*: NIRCam/F322W2, *e*: NIRCam/F444W, and *f*: MIRI/LRS). In general, all pipelines are similarly performant, though our fiducial `Eureka!` analysis outperforms `Pegasus` in the NIRCam/F322W2 observations and outperforms `tshirt` in the MIRI/LRS observations.

Lindsey S. Wiser, Taylor J. Bell, Michael R. Line, Everett Schlawin, Thomas G. Beatty, Luis Welbanks, Thomas P. Greene, Vivien Parmentier, Matthew M. Murphy, Jonathan J. Fortney, Kenny Arnold, Nishil Mehta, Kazumasa Ohno, and Sagnick Mukherjee

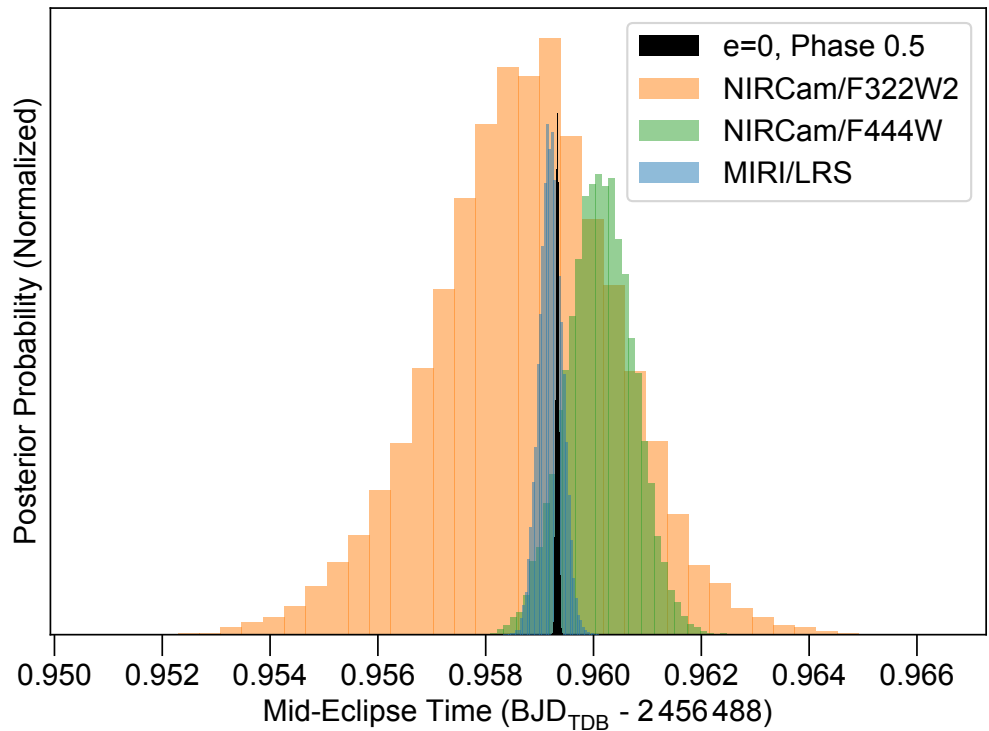

**Fig. S4.** The freely fitted mid-eclipse times from all three observations are shown in orange (NIRCam/F322W2), green (NIRCam/F444W), and blue (MIRI/LRS) contours; priors for all of which were $2{,}456{,}488.959 \pm 0.010$ BJD$_{\text{TDB}}$. All three visits are consistent with each other and with our computed zero-eccentricity mid-eclipse time (shown with a black contour).

Lindsey S. Wiser, Taylor J. Bell, Michael R. Line, Everett Schlawin, Thomas G. Beatty, Luis Welbanks, Thomas P. Greene, Vivien Parmentier, Matthew M. Murphy, Jonathan J. Fortney, Kenny Arnold, Nishil Mehta, Kazumasa Ohno, and Sagnick Mukherjee

**Table S1. The freely fitted mid-eclipse times from all three eclipse observations, compared with the mid-eclipse time that would be predicted from computing the phase=0.5 time from a zero-eccentricity fit to the NIRCam/F322W2 transit of ref. ([1]) with priors from ref. ([2]).**

|  | Mid-Eclipse Time (BJD$_{\mathrm{TDB}}$) | Offset from e=0, phase=0.5 prediction (seconds) |
|---|---|---|
| e=0, phase=0.5 prediction | 2,456,488.959331 $\pm$ 0.000024 | $\equiv 0$ |
| NIRCam/F322W2 | 2,456,488.9588 $\pm$ 0.0017 | -50 +/- 150 |
| NIRCam/F444W | 2,456,488.96013 $\pm$ 0.00062 | 69 +/- 53 |
| MIRI/LRS | 2,456,488.95921 $\pm$ 0.00021 | -11 +/- 18 |

**Lindsey S. Wiser, Taylor J. Bell, Michael R. Line, Everett Schlawin, Thomas G. Beatty, Luis Welbanks, Thomas P. Greene, Vivien Parmentier, Matthew M. Murphy, Jonathan J. Fortney, Kenny Arnold, Nishil Mehta, Kazumasa Ohno, and Sagnick Mukherjee**

7 of 11

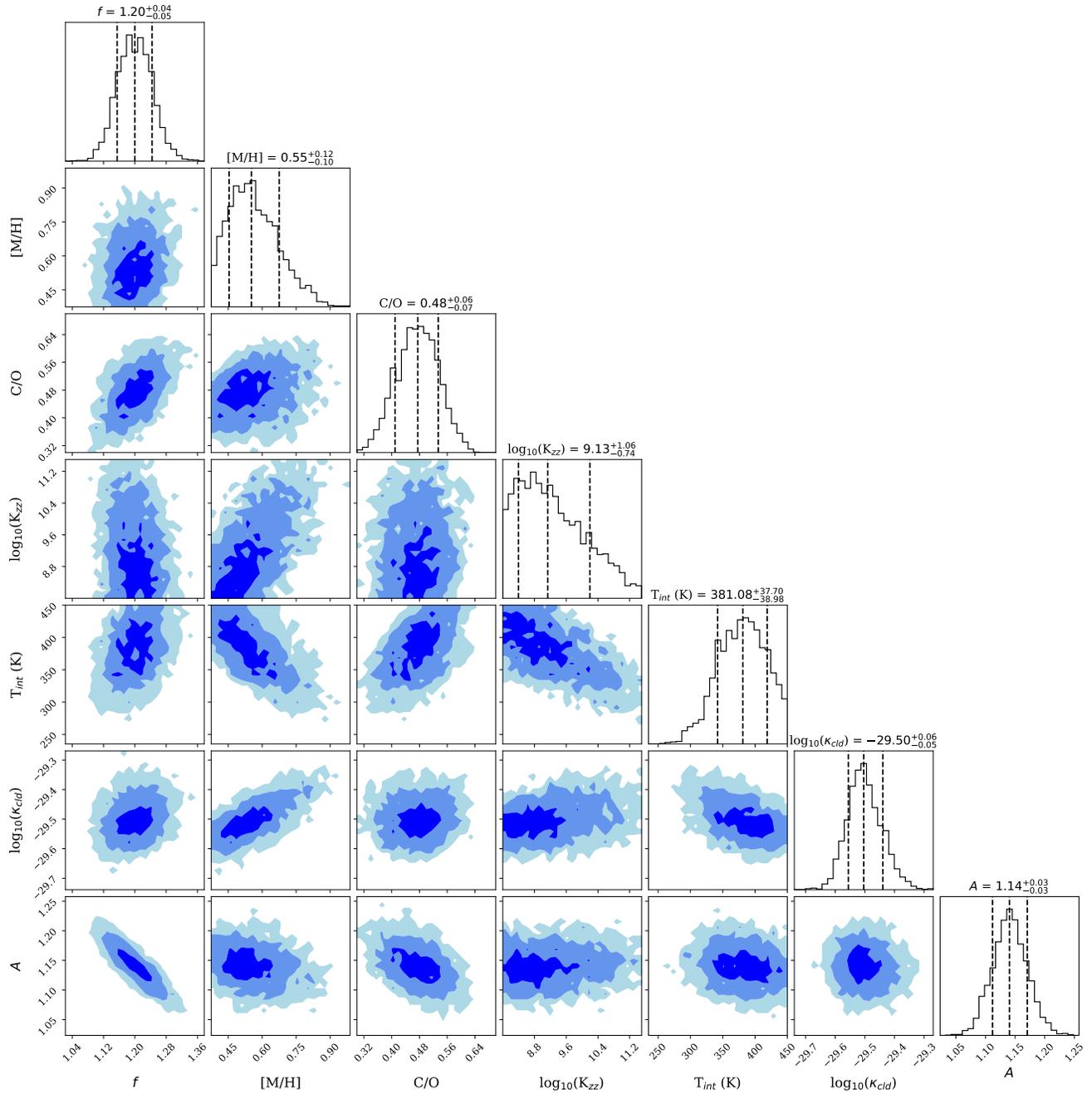

**Fig. S5.** Posterior distribution for the fiducial grid-based retrieval presented in the main manuscript. Parameter estimation medians and $1\sigma$ confidence regions are identified in histogram titles and with vertical lines.

Bradley S. Wiser, Taylor J. Bell, Michael R. Line, Everett Schlawin, Thomas G. Beatty, Luis Welbanks, Thomas P. Greene, Vivien Parmentier, Matthew M. Murphy, Jonathan J. Fortney, Kenny Arnold, Nishil Mehta, Kazumasa Ohno, and Sagnick Mukherjee

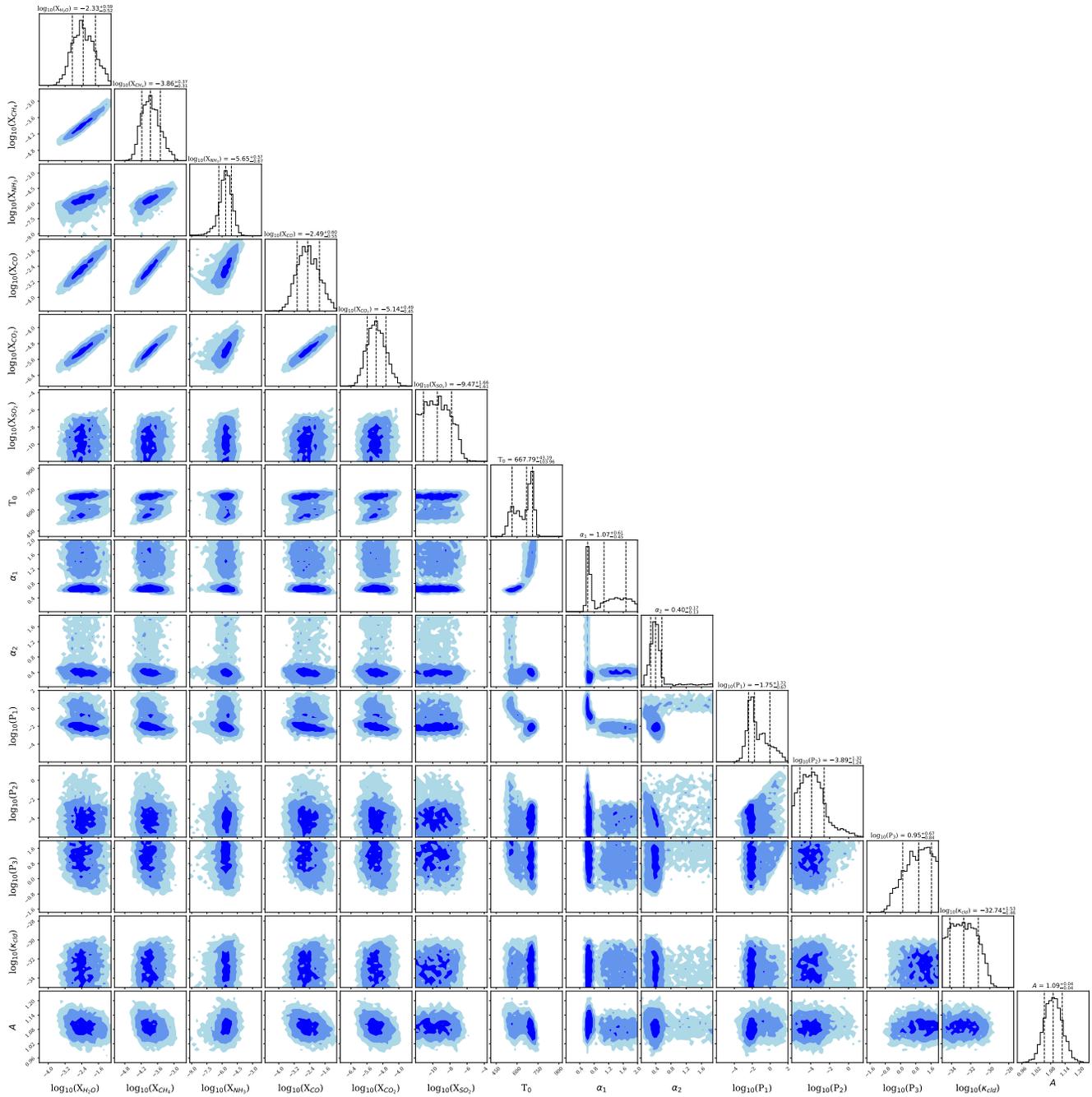

**Fig. S6.** Posterior distribution for the free retrieval. Parameter estimation medians and 1σ confidence regions are identified in histogram titles and with vertical lines.

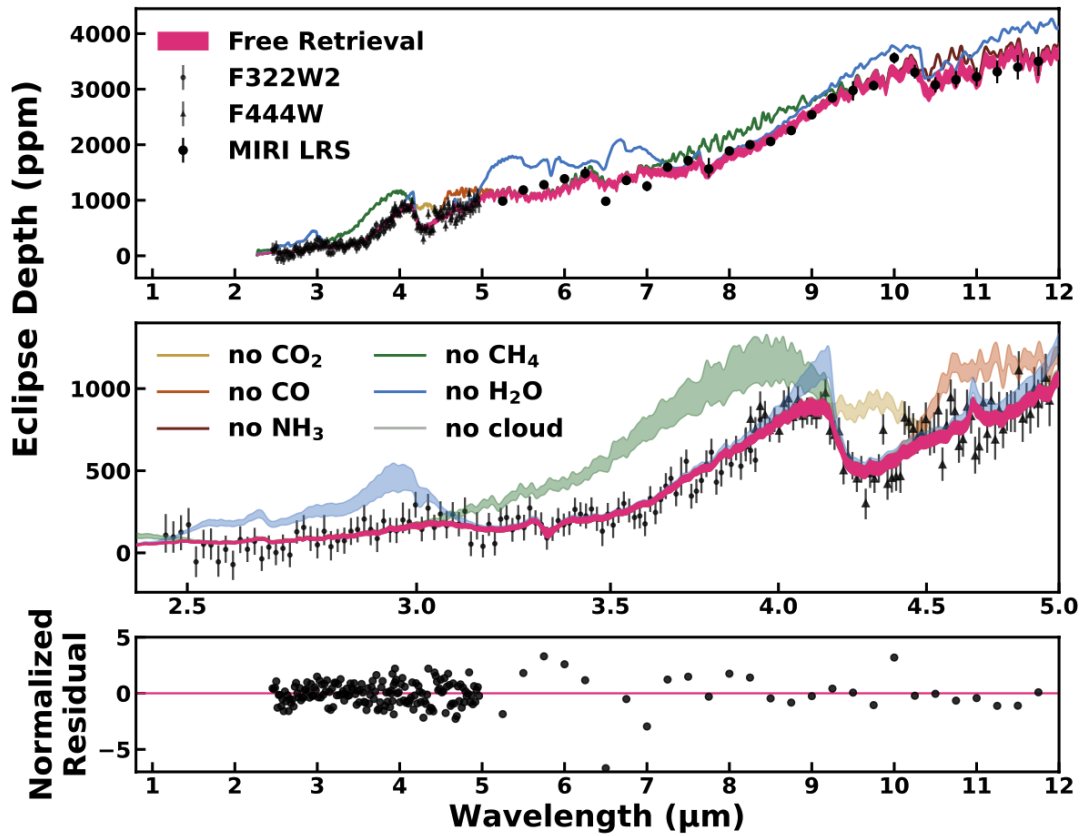

**Fig. S7.** *Top and Middle:* Modeled emission spectra from the free retrieval. The middle plot zooms in on the NIRCam observations. **Pink:** The $2\sigma$ confidence region of the emission spectrum from the free retrieval. **Other colored lines:** The free retrieval spectrum with individual molecules, or the uniform grey cloud, turned "off," illustrating their spectral signatures. On the top, only the median spectrum for each removed molecule/cloud is shown, while in the middle, the $2\sigma$ confidence region is plotted. Notably, as compared to the grid-based retrieval spectrum plotted in the main paper, the free retrieval spectrum shows no significant impact from a grey cloud. *Bottom:* Normalized residuals ([data-model]/error) comparing the best-fit spectrum to the observations.

Lindsey S. Wiser, Taylor J. Bell, Michael R. Line, Everett Schlawin, Thomas G. Beatty, Luis Welbanks, Thomas P. Greene, Vivien Parmentier, Matthew M. Murphy, Jonathan J. Fortney, Kenny Arnold, Nishil Mehta, Kazumasa Ohno, and Sagnick Mukherjee

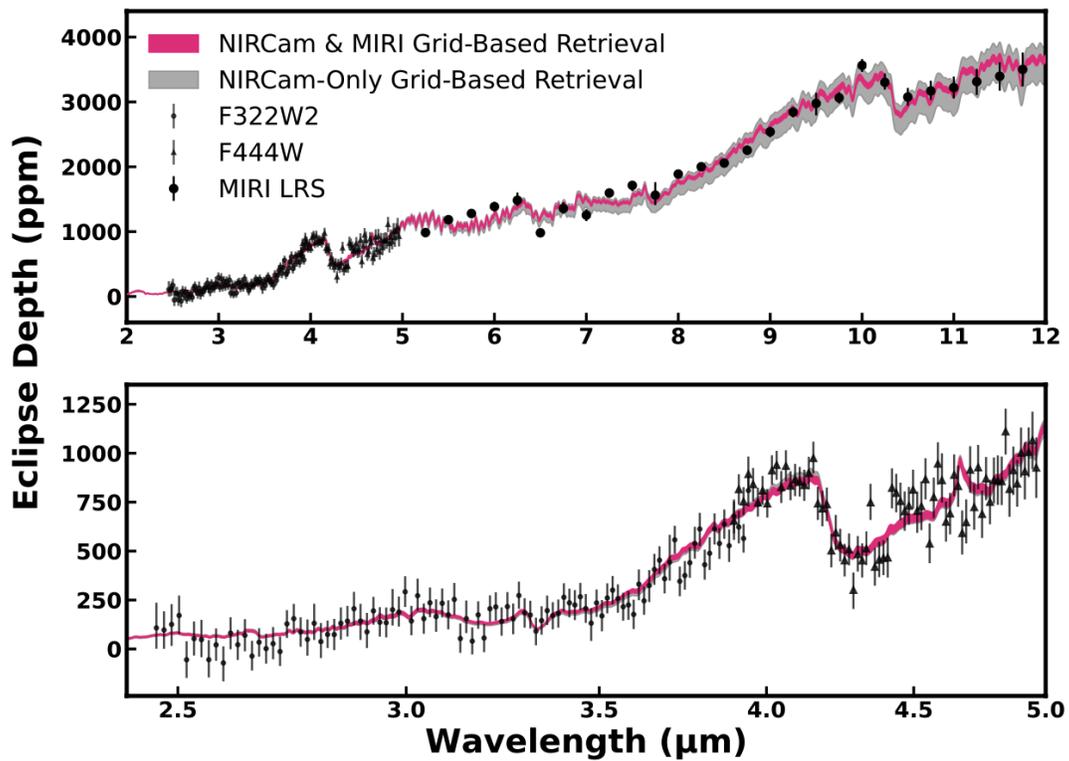

**Fig. S8.** Modeled emission spectra from the fiducial grid-based retrieval to all NIRCam and MIRI observations and a grid-based retrieval to only NIRCam F322W2 and F444W. The bottom plot zooms in on the NIRCam observations. 2$\sigma$ confidence regions are shaded for both retrievals.

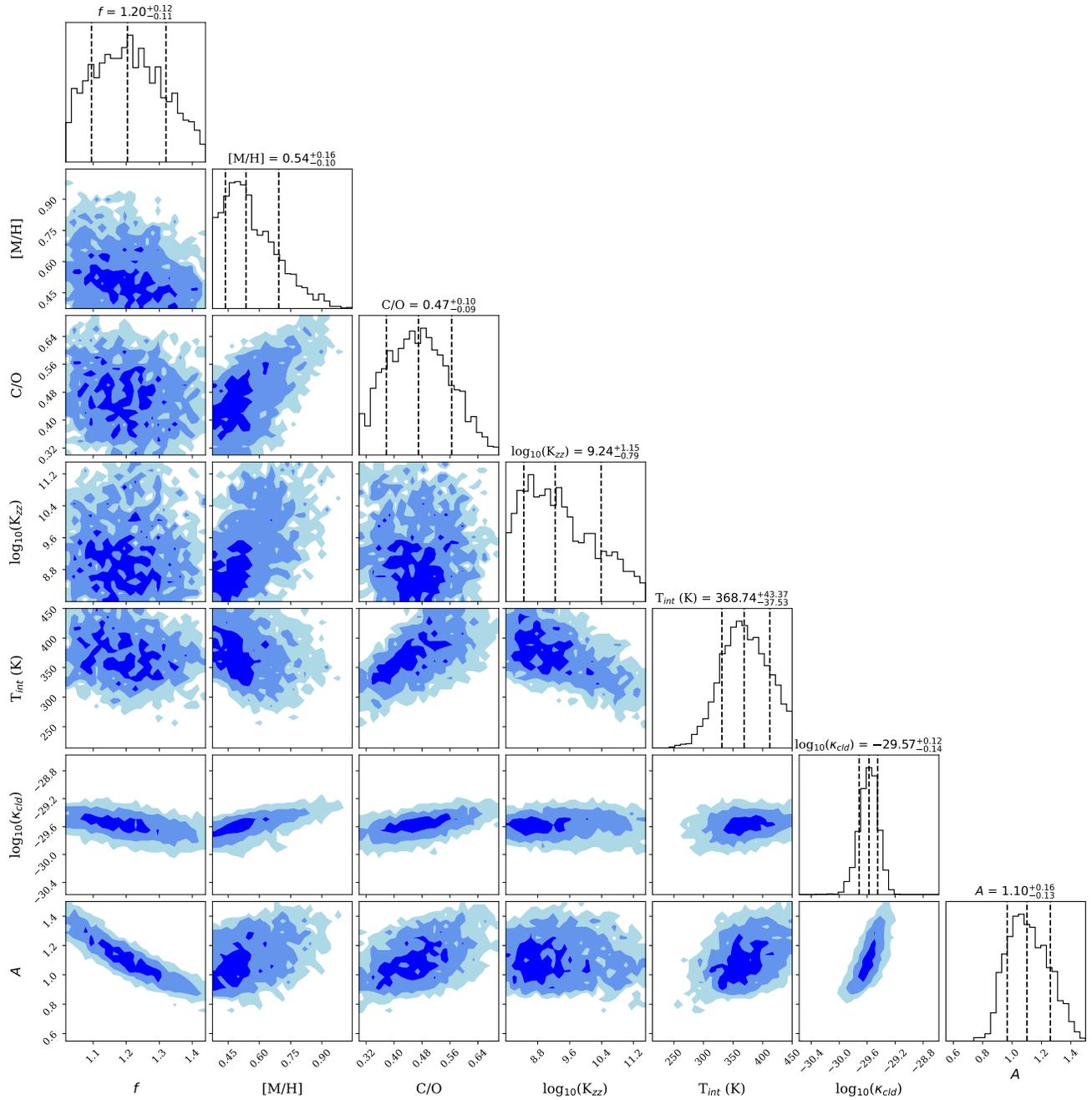

**Fig. S9.** Posterior distribution for a grid-based retrieval to only NIRCam F322W2 and F444W. Parameter estimation medians and $1\sigma$ confidence regions are identified in histogram titles and with vertical lines.

Lindsey S. Wiser, Taylor J. Bell, Michael R. Line, Everett Schlawin, Thomas G. Beatty, Luis Welbanks, Thomas P. Greene, Vivien Parmentier, Matthew M. Murphy, Jonathan J. Fortney, Kenny Arnold, Nishil Mehta, Kazumasa Ohno, and Sagnick Mukherjee

**Table S2.** Fit statistics for our grid-based retrieval and free retrieval presented in the main manuscript, as well as statistics for a grid-based retrieval to only the NIRCam observations. $\chi^2$ p-values $< 0.05$ for the fiducial retrievals indicate low confidence in the data-model fit, however $> 0.5$ Kolmogorov-Smirnov (K-S) p-values indicate high confidence that the data-model residuals are drawn from a normal distribution, i.e. that the residuals may result from random measurement error. Both the $\chi^2$ and K-S tests indicate confidence in the NIRCam only retrieval result.

|  | Grid, Fiducial | Free, Fiducial | Grid, NIRCam Only |
|---|---|---|---|
| $N_{data}$ | 199 | 199 | 172 |
| $N_{parameters}$ | 7 | 14 | 7 |
| $N_{DOF}$ | 192 | 185 | 165 |
| $\chi^2$ | 265.15 | 251.91 | 157.05 |
| $\chi^2/N_{data}$ | 1.33 | 1.27 | 0.91 |
| $\chi^2/N_{DOF}$ | 1.38 | 1.36 | 0.95 |
| $\chi^2$ p-value | $4 \times 10^{-4}$ | $8 \times 10^{-4}$ | 0.66 |
| K-S Statistic | 0.04 | 0.04 | 0.04 |
| K-S p-value | 0.89 | 0.86 | 0.97 |

Lindsey S. Wiser, Taylor J. Bell, Michael R. Line, Everett Schlawin, Thomas G. Beatty, Luis Welbanks, Thomas P. Greene, Vivien Parmentier, Matthew M. Murphy, Jonathan J. Fortney, Kenny Arnold, Nishil Mehta, Kazumasa Ohno, and Sagnick Mukherjee



\end{document}